\documentclass[a4paper,11pt]{article}
\pdfoutput=1 
\usepackage{tabls}
\usepackage{color}
\usepackage{jcappub} 
\usepackage{comment}
\usepackage[T1]{fontenc} 
\usepackage{dsfont}
\title{Generic inference of inflation models by non-Gaussianity and primordial power spectrum reconstruction}
\author[a,1]{Sebastian Dorn,\note{Corresponding author.}}
\author[b]{Erandy Ramirez,}
\author[c]{Kerstin E.\ Kunze,}
\author[d]{Stefan Hofmann,}
\author[a,e]{and Torsten A.\ En{\ss}lin}
\affiliation[a]{Max-Planck-Institut f\"ur Astrophysik,\\ Karl-Schwarzschild-Str.~1, D-85748 Garching, Germany}
\affiliation[b]{Instituto de Ciencias Nucleares, UNAM A. Postal 70-543, Mexico D.F. 04510, Mexico}
\affiliation[c]{Departamento de F\'\i sica Fundamental and IUFFyM, Universidad de Salamanca,\\ Plaza de la Merced s/n, 37008 Salamanca, Spain}
\affiliation[d]{Arnold Sommerfeld Center for Theoretical Physics, Ludwigs-Maximilians-Universit\"at M\"unchen,\\ Theresienstra{\ss}e 37, D-80333 Munich, Germany}
\affiliation[e]{Ludwigs-Maximilians-Universit\"at M\"unchen,\\ Geschwister-Scholl-Platz 1, D-80539 Munich, Germany}
\emailAdd{sdorn@map-garching.mpg.de}

\abstract{We present a generic inference method for inflation models from observational data by the usage of higher-order statistics of the curvature perturbation on uniform density hypersurfaces. This method is based on the calculation of the posterior for the primordial non-Gaussianity parameters $f_\text{NL}$ and $g_\text{NL}$, which in general depend on specific parameters of inflation and reheating models, and enables to discriminate among the still viable inflation models. To keep analyticity as far as possible to dispense with numerically expensive sampling techniques a saddle-point approximation is introduced, whose precision is validated for a numerical toy example. The mathematical formulation is done in a generic way so that the approach remains applicable to cosmic microwave background data as well as to large scale structure data. Additionally, we review a few currently interesting inflation models and present numerical toy examples thereof in two and three dimensions to demonstrate the efficiency of the higher-order statistics method. A second quantity of interest is the primordial power spectrum. Here, we present two Bayesian methods to infer it from observational data, the so called critical filter and an extension thereof with smoothness prior, both allowing for a non-parametric spectrum reconstruction. These methods are able to reconstruct the spectra of the observed perturbations and the primordial ones of curvature perturbation even in case of non-Gaussianity and partial sky coverage. We argue that observables like $T-$ and $B-$modes permit to measure both spectra. This also allows to infer the level of non-Gaussianity generated since inflation.}
\keywords{Cosmic Inflation -- Non-Gaussianity -- Cosmic Microwave Background -- Primordial Power Spectrum -- Large Scale Structure -- Bayesian Inference Method -- Information Field Theory -- Posterior Validation.}

\begin{document}
\excludecomment{commenta}

\maketitle
\flushbottom
\section{Introduction}
\subsection{Motivation}
By precision measurements of the cosmic microwave background (CMB) \cite{2012arXiv1212.5225B,2013arXiv1303.5076P} it has become possible to determine the exact statistics of its temperature anisotropies. These anisotropies are strongly connected to the curvature perturbations on uniform density hypersurfaces $\zeta$, predicted by inflationary models, with the result that the zoo of models can be constrained by exploiting observational data, e.g., by the usage of Gaussian statistics \cite{2006JCAP...08..009M,2013arXiv1303.3787M,2013arXiv1312.2347R,2013arXiv1312.3529M,2013arXiv1303.5082P}. The viable models that are compatible with current \textit{Planck} constraints on primordial non-Gaussianity, often represented by the $f_\text{NL}$ parameter, are given by $-3.1 \leq f_\text{NL} \leq 8.5~(68\%$C.L. statistical$)$ \cite{2013arXiv1303.5084P} for the local type of non-Gaussianity. In particular, a value of  $|f_\text{NL}|\propto \mathcal{O}(1)$ is in agreement with the data. Such a low value of non-Gaussianity opens the possibility to include the effect of primordial non-Gaussianity when performing routine cosmological parameter estimates in order to maximally exploit the data, since it permits approximations which prevent the computations from becoming numerically too expensive. The contributions from higher-order statistics can in many cases (see Sec.\ \ref{sec:special_models}) be parametrized by the local non-Gaussianity parameter $f_\text{NL}$ and $g_\text{NL}$ \cite{2006PhRvD..74j3003S},
\begin{equation}
\label{deffnl}
\zeta = \zeta_1 + \frac{3}{5}f_\text{NL}\zeta_1^2 +\frac{9}{25}g_\text{NL}\zeta_1^3 +\mathcal{O}(\zeta_1^4),
\end{equation}
where $\zeta_1$ is the Gaussian curvature perturbation. $f_\text{NL}$ contributes to the bi- and trispectrum, while $g_\text{NL}$ contributes only to the trispectrum of the curvature perturbation.

As things turned out, there are inflation models among the ones, which are favored by current data, e.g., stated in Ref.\ \cite{2013arXiv1312.3529M} (AI, BI, ESI, HI, LI, MHI, RGI, SBI, SFI )\footnote{Terminology according to Ref.\ \cite{2006JCAP...08..009M}: AI = Arctan Inflation, BI = Brane Inflation, ESI = Exponential SUSY Inflation, HI = Higgs Inflation, LI = Loop Inflation, MHI = Mutated Hilltop Inflation, RGI = Radiation Gauge Inflation, SBI = Supergravity Brane Inflation, SFI = Small Field Inflation.} or Ref.\ \cite{2013arXiv1303.5082P}, predicting values of $|f_\text{NL}|\propto\mathcal{O}(1)$ and distinctly deviate from $g_\text{NL}=0$ if the possibility of non-Gaussianity is taken into account. It is crucial to realize that it is less likely for (at least) two disjunct inflation models to predict the same combination $(f_\text{NL},g_\text{NL})$ than only the same value of $f_\text{NL}$ or $g_\text{NL}$. In other words, if we would be able to infer these two non-Gaussianity parameters simultaneously from CMB or large scale structure (LSS) data, we had a powerful tool to distinguish between the remaining inflation models. This requires to derive a posterior probability density function (pdf) for $(f_\text{NL},g_\text{NL})$ within a Bayesian framework. How this can be done analytically is presented in the first part of this paper. Additionally, we show how this method can be recast to infer parameters specific to inflationary models, e.g., shape parameters of inflationary potentials, or the presence of an additional bosonic field, directly from data. We also provide a validation of our approach to show its precision despite using an approximation. 

The second quantity of interest here is the primordial power spectrum, $P_\zeta(k)$ or $P_{\zeta_1}(k)$, in particular due to its constraining character with respect to inflationary scenarios. The \textit{Planck} collaboration might have seen some features within the primordial power spectrum which in turn would indicate non-linear physics and thus could point to inflation models beyond single-field slow-roll inflation \cite{2013arXiv1303.5082P}. Additionally, these types of deviations are well motivated by, e.g., implications of the recent BICEP2 data \cite{2014arXiv1403.3985B,2014arXiv1403.7786H,2014arXiv1404.3690H,2014arXiv1405.2012H}, or special features of the inflaton potential \cite{2014arXiv1404.2985E,2014arXiv1404.6093M}. However, for the detection of such features one has to appropriately reconstruct the power spectrum from observational data. For this purpose we suggest two non-parametric spectral inference methods in Sec.\ \ref{sec:pps}. 

\subsection{Previous Bayesian work}
The majority of publications \cite{2007JCAP...03..019C,2011PhRvD..84f3013S,2010ApJ...724.1262E,2010A&A...513A..59E,2009PhRvD..80j5005E,PhysRevD.87.063003, KSW}, which are dealing with Bayesian reconstructions of non-Gaussian quantities from CMB have their focus only on estimators or the pdf of the $f_\text{NL}$ or $g_\text{NL}$ parameter. They usually require computationally expensive calculations like Monte Carlo sampling except for some, e.g.\ Refs.\ \cite{paper1, 2013JCAP...06..023V}, which derive analytic expressions by performing approximations.

High precision CMB measurements of the \textit{WMAP} and \textit{Planck} satellites have opened a new window to the physics of the early Universe and have thus improved the constraints on some parameters of non-Gaussianity \cite{2013arXiv1303.5084P} and on many inflation models \cite{2013arXiv1303.5082P} based on the two-point function, but have not connected the inflationary parameters directly to higher-order statistics. 
A way of direct inference of single-field slow-roll inflation models from CMB data of the \textit{Planck} satellite was recently presented by Refs.\ \cite{2013arXiv1303.3787M,2013arXiv1312.2347R,2013arXiv1312.3529M}. Here, the CMB power spectrum was analyzed already ruling out a huge amount of inflation models. We, however,  go beyond Gaussian and three-point statistics to achieve tighter constraints on reasonable, not necessarily single-field slow-roll inflation models given the \textit{Planck} and future data.

An independent cross-check of CMB results is the analysis of the LSS data. Current results for non-Gaussianity values, 
e.g.\ Refs. \cite{2013arXiv1303.1349G,2013arXiv1309.5381A,2008PhRvD..77l3514D,2008JCAP...08..031S,2012MNRAS.422.2854G,2011JCAP...08..033X,2013MNRAS.428.1116R} and 
forwarding references thereof, are consistent with CMB constraints. 
Thus, the LSS provides also a natural data set to infer inflation models. The inference approach presented in 
this paper is in principle able to deal with this type of data sets as well (see Sec.\ 
\ref{sec:validation}).

According to the reconstruction of the primordial power spectrum, there exist a huge amount of approaches and an overview of the literature can be found in section 7 of Ref.\ \cite{2013arXiv1303.5082P} and in Ref.\ \cite{ 2014arXiv1402.1983P}. Within this work we exclusively focus on the approach of Refs.\ \cite{2011PhRvD..83j5014E} and \cite{2013PhRvE..87c2136O}, which developed approximative, but inexpensive Bayesian inference schemes for spectra within the framework of information field theory.

For a brief review on inferring primordial non-Gaussianities in the CMB beyond Bayesian techniques (e.g., bi- and trispectrum estimators, Minkowski Functionals, wavelets, needlets, etc.) we want to point to Ref.\ \cite{2010AdAst2010E..71Y} and forwarding references thereof.

\subsection{Structure of the work}
The remainder of this work is organized as follows. In Sec.\ \ref{sec:ift} we describe the considered data model and introduce the generic method of inferring inflation models postulating $f_\text{NL}$, $g_\text{NL}$. In Sec.\ \ref{sec:special_models} we review a few inflation models that are not ruled out by current \textit{Planck} data and quote corresponding expressions for $f_\text{NL}$ and $g_\text{NL}$. Additionally, we show where the specific models are localized in the $f_\text{NL}$-$g_\text{NL}$-plane. The Bayesian posterior for special inflationary parameters is shown in Sec.\ \ref{sec:validation} as well as a numerical implementation (toy case) of the also pedagogically important curvaton scenario in the Sachs-Wolfe limit and its validation by the Diagnostics of Insufficiencies of Posterior distribution (DIP) test \cite{paper2}. In Sec. \ref{sec:pps} we introduce a method to reconstruct the primordial power spectrum of $\zeta$ and $\zeta_1$. We summarize our findings in Sec.\ \ref{sec:conclusion}.

Being at the interface of statistical analysis and physical cosmology, it seems appropriate to guide the reader by giving some reading instructions. For a reader who is rather interested in the statistical analysis, i.e.\ how to infer (inflationary) parameters from CMB data and how to reconstruct a power spectrum in a non-parametric way in general, paragraphs starting with symbol $\blacktriangleright$ and ending with symbol $\blacktriangleleft$ might be skipped. For a reader rather interested in physical cosmology these symbols might mark paragraphs of special interest.
\section{Generic inference of inflation models postulating $f_\text{NL}$, $g_\text{NL}$}\label{sec:ift}
In order to decide which inflation model is favored by current CMB or LSS data one should use as much information as possible during the inference process without becoming numerically too expensive. This implies, in particular, to aim for information sensitive to non-Gaussian statistics. Usually, this leads to non-trivial phase space integrals which cannot be performed analytically and require numerically expensive techniques like Monte Carlo sampling. Within this section, however, we show how to set up a fully analytic posterior for the scalar, local non-Gaussianity parameters $f_\text{NL}(p)$ and $g_\text{NL}(p)$, which in general depend on inflation or reheating model specific parameters, $p=(p_1,\dots,p_u)^T\in \mathds{R}^u,~u\in \mathds{N}$. In turn, this also enables to calculate analytically the posterior pdf for the model specific parameters $p$, which encode, e.g., the particular shape of an inflation model or the density fraction of an additional bosonic field (see Sec.\ \ref{sec:special_models}). To keep this analyticity and simultaneously avoid numerically expensive sampling techniques we introduce a saddle-point approximation in the actual section, whose sufficiency is validated in Sec.\ \ref{sec:valid}. 

For reasons of clarity and comprehensibility we drop the $p$-dependency in our notation within this section. For the same reason we focus on global values of $f_\text{NL}$ and $g_\text{NL}$ although the formalism described below is generic and can deal with spatially varying non-Gaussianity parameters as shown in Ref.\ \cite{2009PhRvD..80j5005E}.  

\subsection{Data model}\label{sec:data}
To infer physical quantities from data we have to agree on a particular data model. Following the logic of information field theory \cite{2009PhRvD..80j5005E}, a CMB observation is represented by a discrete data tuple $d=(d_1,\dots,d_m)^T\in \mathds{R}^m,~m\in\mathds{N}$, composed of uncorrelated Gaussian noise $n=(n_1,\dots,n_m)^T\in\mathds{R}^m$ and a linear response operation $R$ acting on the, in general, non-Gaussian comoving curvature perturbations $\zeta$, a continuous physical field over the Riemannian manifold $\mathcal{U}$,
\begin{equation}
\label{data}
d=\frac{\delta T}{T_\text{CMB}} = R \zeta + n = R\left(\zeta_1 + \frac{3}{5}f_\text{NL}\zeta_1^2 +\frac{9}{25}g_\text{NL}\zeta_1^3 +\mathcal{O}\left(\zeta_1^4\right)\right) +n,
\end{equation}
with the Gaussian curvature perturbations $\zeta_1$. The pdfs of $\zeta_1$ and $n$ are given by $P(\zeta_1)=\mathcal{G}(\zeta_1,\Xi)$ with covariance $\Xi = \left\langle\zeta_1\zeta_1^\dag\right\rangle_{(\zeta_1|\Xi)}$, and $P(n) = \mathcal{G}(n,N)$ with covariance $N$, respectively. Here, we use the notation
\begin{equation}
\left\langle ~.~ \right\rangle_{P(a)} = \left\langle ~.~ \right\rangle_{(a|A)} \equiv  \int \mathcal{D}a ~.~P(a|A),
\end{equation}
and
\begin{equation}
\mathcal{G}(a,A)\equiv  |2\pi A|^{-1/2}~\exp\left(-\frac{1}{2}a^\dag A^{-1} a \right),   
\end{equation}
where $\dag$ denotes a transposition and complex conjugation, $*$, and $a^\dag b \equiv \int_{\mathcal{U}}d^{\mathrm{d}}x ~a^*(x)b(x)$ with $\mathrm{d}\equiv \dim\mathcal{U}$ defining the inner product on the fields $a,~b$. The comoving curvature perturbation $\zeta$ on uniform density hypersurfaces, which is a conserved quantity outside the horizon\footnote{Note that this is true in the standard 
$\Lambda CDM$ model. However, if there are sources of anisotropic stress before neutrino decoupling, such as, e.g., in the case of primordial magnetic fields, then $\zeta$ is no longer a constant on superhorizon scales \cite{2010JCAP...02..018K,2010PhRvD..81d3517S}.} \cite{2013PhRvD..87b3530S}, is the seed of the structure growth during the evolution of the Universe and its statistics are precisely predicted by inflation models. Therefore, $\zeta$ is directly related to inflationary parameters, $p$. If the statistics of $\zeta$, predicted by inflation scenarios, are non-Gaussian, the dependence on $p$ can often be absorbed in the non-Gaussianity parameters $f_\text{NL}(p)$ and $g_\text{NL}(p)$. The linear response $R$ in Eq.\ (\ref{data}) transfers the curvature perturbations into temperature deviations, $\delta T$, and contains all instrumental and measurement effects, i.e.\ $R$ represents the radiation transfer function. In this way the data is directly related to the initial Gaussian curvature perturbation $\zeta_1$ or to inflationary parameters $p$ and we can set up the inference scheme.

\subsection{Posterior derivation}\label{sec:postder}
We derive the posterior by following Ref.\ \cite{paper1}, i.e.\ we first calculate the pdf for the Gaussian curvature perturbation $\zeta_1$ given the non-Gaussianity parameters and data via Bayes theorem \cite{Bayes01011763},
\begin{equation}
\begin{split}
\label{Bayes}
P(\zeta_1|d,f_\text{NL},g_\text{NL}) =&~ \frac{P(\zeta_1, d,f_\text{NL},g_\text{NL})}{P(d,f_\text{NL},g_\text{NL})} \\ =&~\frac{P(d|\zeta_1,f_\text{NL},g_\text{NL})P(\zeta_1|f_\text{NL},g_\text{NL})}{P(d|f_\text{NL},g_\text{NL})} \equiv \frac{1}{\mathcal{Z}}e^{-H(\zeta_1,d|f_\text{NL},g_\text{NL})},
\end{split}
\end{equation}
where $H(\zeta_1,d|f_\text{NL},g_\text{NL})\equiv -\ln[P(d|\zeta_1,f_\text{NL},g_\text{NL})P(\zeta_1|f_\text{NL},g_\text{NL})]$ defines the information Hamiltonian and $\mathcal{Z}\equiv P(d|f_\text{NL},g_\text{NL})$ the partition function. Assuming Gaussian noise, $\mathcal{G}(n,N)$,
with $N=\left\langle nn^\dag\right\rangle_{(n|N)}$ denoting the noise covariance matrix and that $f_\text{NL}$ and $g_\text{NL}$ are constant scalars and that all quantities are real, the information Hamiltonian is given by
\begin{equation}
\label{hamgen}
\begin{split}
H&(\zeta_1,d|f_\text{NL},g_\text{NL})\\
 =&~ -\ln\left[ P(d|\zeta_1,f_\text{NL},g_\text{NL})P(\zeta_1|f_\text{NL},g_\text{NL})\right] = -\ln\left[\mathcal{G}\left(d-R\zeta,N\right)\mathcal{G}\left(\zeta_1,\Xi\right) \right]\\
=&~H_0 + \frac{1}{2}\zeta^\dag_1 D^{-1}\zeta_1 - j^\dag\zeta_1 -\frac{3}{5}f_\text{NL}j^\dag\zeta_1^2 -\frac{9}{25}g_\text{NL}j^\dag\zeta_1^3 +\frac{3}{5}f_\text{NL}\zeta_1^\dag M \zeta_1^2\\ 
	&+\frac{9}{25}g_\text{NL}\zeta_1^\dag M \zeta_1^3 +\frac{9}{50}f^2_\text{NL}\left(\zeta_1^\dag\right)^2 M \zeta_1^2 + \frac{27}{125}f_\text{NL}g_\text{NL}\left(\zeta_1^\dag\right)^2 M \zeta_1^3 + \frac{81}{1250}g^2_\text{NL}\left(\zeta_1^\dag\right)^3 M \zeta_1^3.
\end{split}
\end{equation}
Note that some terms of order $\mathcal{O}(\zeta_1^5)$ have already been neglected because we did not state the exact expression for the term proportional to $\mathcal{O}(\zeta_1^4)$ in Eq.\ (\ref{data}). Eq.\ (\ref{hamgen}) contains the abbreviations
\begin{equation}
\begin{split}
D^{-1}=&~ \Xi^{-1} + M,~~ M=R^\dag N^{-1}R,~~ j=R^\dag N^{-1}d,\\
\text{and}~H_0=&~\frac{1}{2}\ln|2\pi \Xi| + \frac{1}{2}\ln|2\pi N| +\frac{1}{2}d^\dag N^{-1}d.
\end{split}
\end{equation}

Now, we are able to determine the posterior for the non-Gaussianity parameters $f_\text{NL}$ and $g_\text{NL}$, which can be calculated by combining Eqs.\ (\ref{Bayes}) and (\ref{hamgen}),
\begin{equation}
\begin{split}
P(f_\text{NL},g_\text{NL}|d) =&~ \frac{P(d|f_\text{NL},g_\text{NL})P(f_\text{NL},g_\text{NL})}{P(d)}
\propto P(f_\text{NL},g_\text{NL})\int\mathcal{D}\zeta_1 P(\zeta_1, d|f_\text{NL},g_\text{NL}) \\
 =&~ P(f_\text{NL},g_\text{NL})\int\mathcal{D}\zeta_1 \exp\left[-H(\zeta_1,d|f_\text{NL},g_\text{NL})\right].
\end{split}
\end{equation}
Due to the fact that the Hamiltonian contains higher orders than $\zeta_1^2$ we cannot perform the path-integration analytically. To circumvent this obstacle we conduct a saddle-point approximation in $\zeta_1$ around $\bar{\zeta_1}\equiv \arg\min\left[H(\zeta_1,d|f_\text{NL},g_\text{NL})\right]$ up to the second order in $\zeta_1$ to be still able to perform the path-integration analytically, cf.\ \cite{paper1}. This Taylor approximation is justified by $|\zeta_1| \propto \mathcal{O}\left(10^{-5}\right)$. For the expansion of the Hamiltonian we need the first and second derivative with respect to $\zeta_1$, given by

\begin{equation}
\label{gradgen}
\begin{split}
0=&~\frac{\delta H(\zeta_1,d|f_\text{NL},g_\text{NL})}{\delta \zeta_1}\bigg\vert_{\zeta_1=\bar{\zeta_1}} \\
=&~ \left(D^{-1}-\frac{6}{5}f_\text{NL} ~\hat{j}\right)\bar{\zeta_1} - j - \frac{27}{25}g_\text{NL} \hat{j}\bar{\zeta_1}^2 +\frac{3}{5}f_\text{NL} \left(M \bar{\zeta_1}^2 +2\bar{\zeta_1} \star M\bar{\zeta_1}\right)  \\
 & +\frac{9}{25}g_\text{NL}\left( M \bar{\zeta_1}^3 +3 \bar{\zeta_1}^2\star M\bar{\zeta_1}\right) + \left(\frac{18}{25}f_\text{NL}^2 \bar{\zeta_1}\star M \bar{\zeta_1}^2\right)\\
 & + \frac{27}{125}f_\text{NL} g_\text{NL}\left(2\bar{\zeta_1} \star M \bar{\zeta_1}^3 + 3\bar{\zeta_1}^2 \star M\bar{\zeta_1}^2\right) + \left(\frac{243}{625}g_\text{NL}^2 \bar{\zeta_1}^2 \star M\bar{\zeta_1}^3\right), 
\end{split}
\end{equation}

\noindent and 

\begin{equation}
\label{hessgen}
\begin{split}
D^{-1}_{d,f_\text{NL},g_\text{NL}}\equiv &~\frac{\delta^2 H(\zeta_1,d|f_\text{NL},g_\text{NL})}{\delta \zeta_1^2}\bigg\vert_{\zeta_1=\bar{\zeta_1}} \\
 =&~ D^{-1} -\frac{6}{5}f_\text{NL}\hat{j} - \frac{54}{25}g_\text{NL} \widehat{\bar{\zeta_1}\star j} +\frac{6}{5}f_\text{NL}\left(2\bar{\zeta_1}\star M + \widehat{M\bar{\zeta_1}}\right)\\
	 &~+\frac{27}{25}g_\text{NL}\left(\widehat{M\bar{\zeta_1}^2} +\left(\bar{\zeta_1}^2\star M + 2\widehat{\bar{\zeta_1}}\widehat{M\bar{\zeta_1}}\right)\right) +\frac{18}{25}f_\text{NL}^2\left(\widehat{M\bar{\zeta_1}^2}+2\bar{\zeta_1}\star M\star \bar{\zeta_1}\right)\\	
&~+\frac{54}{125}f_\text{NL}g_\text{NL}\left(3\widehat{\bar{\zeta_1}}\widehat{ M\bar{\zeta_1}^2} +\widehat{M\bar{\zeta_1}^3} +6\bar{\zeta_1}\star M\star \bar{\zeta_1}^2\right)\\
&~ +\frac{243}{625}g_\text{NL}^2\left(2\widehat{\bar{\zeta_1}}\widehat{ M\bar{\zeta_1}^3} +3\bar{\zeta_1}^2\star M\star \bar{\zeta_1}^2\right),
\end{split}
\end{equation}
where $\star$ denotes a pixel-by-pixel multiplication, e.g., $\zeta^2_x=(\zeta \star \zeta)_x \equiv  \zeta_x \zeta_x$ and a hat over fields denotes the transformation of a field to a diagonal matrix, $\zeta_x \mapsto \zeta_x\delta_{xy} \equiv \hat{\zeta}_{xy}$.

With Eqs.\ (\ref{gradgen}) and (\ref{hessgen}) we are able to perform the saddle-point approximation of the posterior yielding 
\begin{equation}
\label{postgen}
\begin{split}
P(f_\text{NL},g_\text{NL}|d)\propto& ~P(f_\text{NL},g_\text{NL})\int \mathcal{D}\zeta_1 \exp\left[-H(d,\zeta_1|f_\text{NL},g_\text{NL})\right] \\
\approx&~ P(f_\text{NL},g_\text{NL})\int \mathcal{D}(\zeta_1-\bar{\zeta_1})\left|\frac{\delta (\zeta_1-\bar{\zeta_1})}{\delta \zeta_1} \right|^{-1}\\  &~\times \exp\left[-H(d,\bar{\zeta_1}|f_\text{NL},g_\text{NL}) - \frac{1}{2}(\zeta_1 - \bar{\zeta_1})^\dag D^{-1}_{d,f_\text{NL},g_\text{NL}}(\zeta_1 - \bar{\zeta_1})\right]\\
 =&~ |2\pi D_{d,f_\text{NL},g_\text{NL}}|^{\frac{1}{2}}\exp\left[-H(d,\bar{\zeta_1}|f_\text{NL},g_\text{NL})\right]P(f_\text{NL},g_\text{NL}).
\end{split}
\end{equation}
Considering Eq.\ (\ref{postgen}), we are able to calculate analytically the full posterior pdf of the $f_\text{NL}$ and $g_\text{NL}$ parameter without using expensive Monte Carlo sampling techniques. These techniques have been avoided by replacing the joint pdf for data and curvature perturbation, $P(\zeta_1,d|f_\text{NL},g_\text{NL})$, by the Gaussian distribution $\mathcal{G}(\zeta_1 - \bar{\zeta_1}, D_{d,f_\text{NL},g_\text{NL}})$, whose precision is validated for particular inflation models in Sec.\ \ref{sec:validation} as well as in Ref.\ \cite{paper1} by applying the DIP test \cite{paper2}. 

Note that the evaluation of Eq.\ (\ref{postgen}) requires a priori knowledge about the primordial power spectrum, $\Xi$ (see Sec.\ \ref{sec:pps} for a more detailed description). In the realistic case of small non-Gaussianity one might try, for instance, to see what consequences the power-law power spectrum of the \textit{Planck} cosmology \cite{2013arXiv1303.5082P} yields, Eq.\ (\ref{powerdef}), as long as $\Xi =  \left\langle\zeta_1\zeta_1^\dag\right\rangle_{(\zeta_1|\Xi)} \approx \left\langle\zeta\zeta^\dag\right\rangle_{(\zeta|\Xi)}$ holds. In regimes of larger non-Gaussianity, where the last approximation is violated, the primordial power spectrum and the reconstruction of $\zeta_1$ (wherefore we need a priori $\Xi$) have to be inferred simultaneously from the data. For this purpose we introduce an Empirical Bayes method in Sec.\ \ref{sec:pps}.

\section{Special models of inflation}\label{sec:special_models}
$\blacktriangleright$ There is a large number of different inflationary models, so what particular type should one focus on? Fortunately, recently published papers given by Refs.\ \cite{2013arXiv1303.5082P} and \cite{2013arXiv1312.3529M} address this question. The first by mainly pointing out parameter constraints to many representative inflation models as well as a Bayesian model comparison thereof, the second by suggesting to concentrate on nine specific types of single-field slow-roll inflation, which are favored by current \textit{Planck} data. To be more precise, the favors of Ref.\  \cite{2013arXiv1312.3529M} have been determined by calculating the Bayesian evidence and complexity of the models.

The remaining nine models are all single-field slow-roll models and thus are characterized by, e.g., two slow-roll parameters\footnote{Analogously one can use the HFF slow-roll parameter definition as done in Ref.\  \cite{2013arXiv1312.3529M}.}, $\epsilon$ and $\eta$, which are given by
\begin{equation}
\label{SRP}
\epsilon \equiv \frac{1}{2}\left(\frac{M_\text{Pl}V_\phi}{V}\right)^2 ~~\text{and}~~\eta \equiv \frac{M_\text{Pl}^2 V_{\phi\phi}}{V},
\end{equation}
where $V\equiv V(\phi)$ denotes the potential of the inflaton $\phi$, $M_\text{Pl}$ the \textit{Planck} mass, subscript letters represent derivatives, and $\epsilon$ and $\eta$ fulfill the bounds $\epsilon, |\eta| \ll 1$.  For these specific models the non-Gaussianity parameters are usually much smaller than one and can often be written as a function of $\epsilon$ and $\eta$, e.g., for single-field slow-roll inflation models with standard kinetic term and Bunch-Davis vacuum as initial vacuum state the parameter $f_\text{NL}$ is proportional to $\mathcal{O}(\epsilon, \eta)$ (for details see \cite{2003NuPhB.667..119A,2003JHEP...05..013M,2013arXiv1303.5084P}). In particular, the quantitative dependence of the non-Gaussianity parameters on inflationary parameters $p$ can be worked out for every inflation model by conducting cosmological perturbation theory \cite{1984PThPS..78....1K,1992PhR...215..203M} to desired order or by applying the so-called $\delta N$-formalism \cite{1985JETPL..42..152S,1996PThPh..95...71S,1998PThPh..99..763S,2013PhRvD..87b3530S}. This means, by replacing the non-Gaussianity parameters by $(\epsilon,~\eta)$-dependent functions, which again depend on inflation model specific parameters $p$ as clarified in Eq.\ (\ref{SRP}), we are able to infer the slow-roll parameters as well as $p$ directly from data according to Eq.\ (\ref{postgen}). Unfortunately, such a tiny amount of non-Gaussianity is currently expected not to be observable due to other general relativistic effects (for details see \cite{2013arXiv1303.5084P}).

The other case of inflation models with Lagrangians including non-standard kinetic terms leads to non-Gaussianity of equilateral type depending on the so-called sound speed of the inflaton, $c_s$ (= 1 for standard kinetic terms) \cite{2013arXiv1303.5082P}. This type of non-Gaussianity can approximately be described by the parameter $f_\text{NL}^\text{eq}\propto (1-c_s^{-2})$. The sound speed of the inflaton, again, depends on the particular inflation model and its parameters. Unfortunately, this type of non-Gaussianity cannot be expressed in a form similar to Eq.\ (\ref{deffnl}) and one has to go back, for instance, to templates.

There are also multi-field inflation models that are not ruled out yet \cite{2013arXiv1303.5082P}. In particular models where initially isocurvature perturbations are (not necessarily completely) transformed to adiabatic perturbations. Such a transformation requires at least two fields, e.g., as it happens in the curvaton, axion, higgs inflation scenario, and indeed a slight favor to a non-vanishing amount of isocurvature modes might have been seen by \textit{Planck} \cite{2013arXiv1303.5082P}. In this section we focus on such scenarios and review the calculations of the non-Gaussianity parameters for a selection of realistic models without claiming that this selection is the most favored one. We are picking only one representative single-parameter mechanism per inflation model for minimal complexity and simplicity. Note, that the approach of Sec.\ \ref{sec:ift} would also allow to focus on the other inflation scenarios mentioned above. We choose the following scenarios for illustration only.~$\blacktriangleleft$

\subsection{Simplest curvaton model}
$\blacktriangleright$ Here, the simplest curvaton model, taking into account radiation and the curvaton, is considered.
Without interactions perfect fluids have conserved curvature perturbations \cite{2002PhRvD..65l1301B,2006PhRvD..74j3003S}, 
\begin{eqnarray}
\zeta_i=\delta N+\frac{1}{3}\int_{\bar{\rho}_i}^{\rho_i}\frac{d\tilde{\rho}_i}{\tilde{\rho}_i+P_i(\tilde{\rho}_i)}, ~~i\in\{r,\chi\},
\end{eqnarray}
with $r$ denoting radiation, $\chi$ the curvaton, $\delta N$ the perturbation of the number of $e$-folds $N$ during inflation, $\rho$ the particle specific density, and $P$ the respective pressure. Barred quantities refer to homogeneous background values. Assuming the curvaton decays on a uniform total density hypersurface determined by $H=\Gamma$, where $H=\frac{\dot{a}}{a}$ is the Hubble parameter, $a$ the cosmological scale factor, and $\Gamma$ the decay rate of the curvaton (assumed to be constant). Then on this hypersurface
\begin{eqnarray}
\rho_r(t_\text{decay},\vec{x})+\rho_{\chi}(t_\text{decay},\vec{x})=\bar{\rho}(t_\text{decay}).
\label{e1}
\end{eqnarray}
However, the local curvaton and radiation densities on this decay surface will be inhomogeneous, with $\zeta=\delta N$,
\begin{eqnarray}
\zeta_r&=&\zeta+\frac{1}{4}\ln\left(\frac{\rho_r}{\bar{\rho}_r}\right)\Rightarrow \rho_r=\bar{\rho}_r~e^{4(\zeta_r-\zeta)},
\label{e2}\\
\zeta_{\chi}&=&\zeta+\frac{1}{3}\ln\left(\frac{\rho_{\chi}}{\bar{\rho}_{\chi}}\right)\Rightarrow \rho_{\chi}=\bar{\rho}_{\chi}~e^{3(\zeta_{\chi}-\zeta)}.
\label{e3}
\end{eqnarray}
Here it was used that  once the curvaton starts oscillating it effectively behaves as a non-relativistic perfect fluid, $\rho_{\chi}\propto a^{-3}$.
Using Eqs.\ (\ref{e2}) and (\ref{e3}) in Eq.\ (\ref{e1}) leads to \cite{2006PhRvD..74j3003S}
\begin{eqnarray}
\Omega_{\chi,\text{decay}}~e^{3(\zeta_{\chi}-\zeta)}+\left(1-\Omega_{\chi,\text{decay}}\right)e^{4(\zeta_r-\zeta)}=1,
\label{e4}
\end{eqnarray} 
where 
\begin{eqnarray}
\Omega_{\chi,\text{decay}}\equiv\frac{\bar{\rho}_{\chi}}{\bar{\rho}_{\chi}+\bar{\rho}_{r}}.
\end{eqnarray}
This equation will now be expanded order by order in $\zeta$.
Following Ref.\ \cite{2006PhRvD..74j3003S} the simplest case is considered in which any perturbation in the radiation fluid is neglected, due to, say an inflationary curvature perturbation. Hence $\zeta_r=0$.

To first order Eq.\ (\ref{e4}) reads \cite{2006PhRvD..74j3003S}
\begin{eqnarray}
\label{expandeq}
4\left(1-\Omega_{\chi, \text{decay}}\right)\zeta_1=3~\Omega_{\chi, \text{decay}}\left(\zeta_{\chi_1}-\zeta_1\right),
\end{eqnarray}
where subscript $1$ denotes the first order expansion so that 
\begin{eqnarray}
\zeta_1=r\zeta_{\chi_1},
\end{eqnarray}
where 
\begin{eqnarray}
r\equiv\frac{3\Omega_{\chi,\text{decay}}}{4-\Omega_{\chi,\text{decay}}}=\left.\frac{3\bar{\rho}_{\chi}}{3\bar{\rho}_{\chi}+4\bar{\rho}_r}\right|_{t_\text{decay}} \in [0,1].
\end{eqnarray}
For $r \approx 1$, i.e.\ the curvaton is highly dominant, the curvature perturbations are purely adiabatic whereas for $r\ll1$ not all isocurvature modes have converted to adiabatic ones. Assuming that the curvaton energy density is determined by a simple quadratic potential,
\begin{eqnarray}
\rho_{\chi}=\frac{1}{2}m^2\chi^2,
\end{eqnarray}
and assuming it is a weakly coupled field during inflation so that its quantum fluctuations induce a classical Gaussian random field 
after horizon exit on superhorizon scales, then
\begin{eqnarray}
\chi_*=\bar{\chi}_*+\delta_1\chi_*,
\end{eqnarray}
where the * indicates the time of horizon exit and 1 in the perturbation emphasizes the linear perturbation.
Moreover, $\delta_1\chi_*$ is a Gaussian random field with 2-point correlation function in $k$-space,
\begin{eqnarray}
\langle\delta_1\chi_{*,\vec{k}}\, \delta_1\chi_{*,\vec{k}'}^{\dag}\rangle=\frac{2\pi^2}{k^3}\left(\frac{H}{2\pi}\right)^2\delta_{\vec{k}\vec{k}'},
\end{eqnarray}
where $H\simeq const.$ during inflation.
Now there could be a nonlinear evolution of  $\chi$ on superhorizon scales after horizon exit up to the beginning of the curvaton oscillations and subsequent decay during the radiation dominated era.
In Ref.\ \cite{2006PhRvD..74j3003S} this is taken into account by introducing a function $g(\chi)$ such that during the curvaton oscillations the value of the curvaton field is given by
\begin{eqnarray}
\chi=g(\chi_*).
\end{eqnarray}
Hence $\bar{\rho}_{\chi}=\frac{1}{2}m^2\bar{g}^2$ and \cite{2006PhRvD..74j3003S}
\begin{eqnarray}
\zeta_{\chi_1}=\frac{2}{3}\frac{\delta_1\chi}{\bar{\chi}}=\frac{2}{3}\left.\frac{g'}{g}\right|_{\chi=\chi_*}\delta_1\chi_*.
\end{eqnarray}
In real space the nonlinearity parameters $f_\text{NL}$ and $g_\text{NL}$ are defined by (e.g. \cite{2006PhRvD..74j3003S})
\begin{eqnarray}
\zeta(t,\vec{x})=\zeta_1(t,\vec{x})+\frac{3}{5}f_\text{NL}\zeta_1(t,\vec{x})^2+\frac{9}{25}g_\text{NL}\zeta_1(t,\vec{x})^3+{\cal O}(\zeta_1^4).
\label{zeta}
\end{eqnarray}
The Bardeen potential $\Phi_\text{md}$ on large scales in the matter dominated era is related to 
$\zeta_1$ by $\Phi_\text{md}=\frac{3}{5}\zeta_1$ so that
\begin{eqnarray}
\frac{3}{5}\zeta=\Phi_\text{md}+f_\text{NL}\Phi^2_\text{md}+g_\text{NL}\Phi_\text{md}^3.
\end{eqnarray}
In Ref.\ \cite{2006PhRvD..74j3003S} $f_\text{NL}$ and $g_\text{NL}$ are calculated by expanding Eq.\ (\ref{expandeq}) respectively up to second and third order, e.g.,
\begin{eqnarray}
f_\text{NL}=\frac{5}{4r}\left(1+\frac{gg''}{g^{' 2}}\right)-\frac{5}{3}-\frac{5r}{6},
\end{eqnarray}
and $g_\text{NL}$ can be found in Ref.\ \cite{2006PhRvD..74j3003S}.

\begin{figure}[ht]
\begin{center}
\includegraphics[width=.7\textwidth]{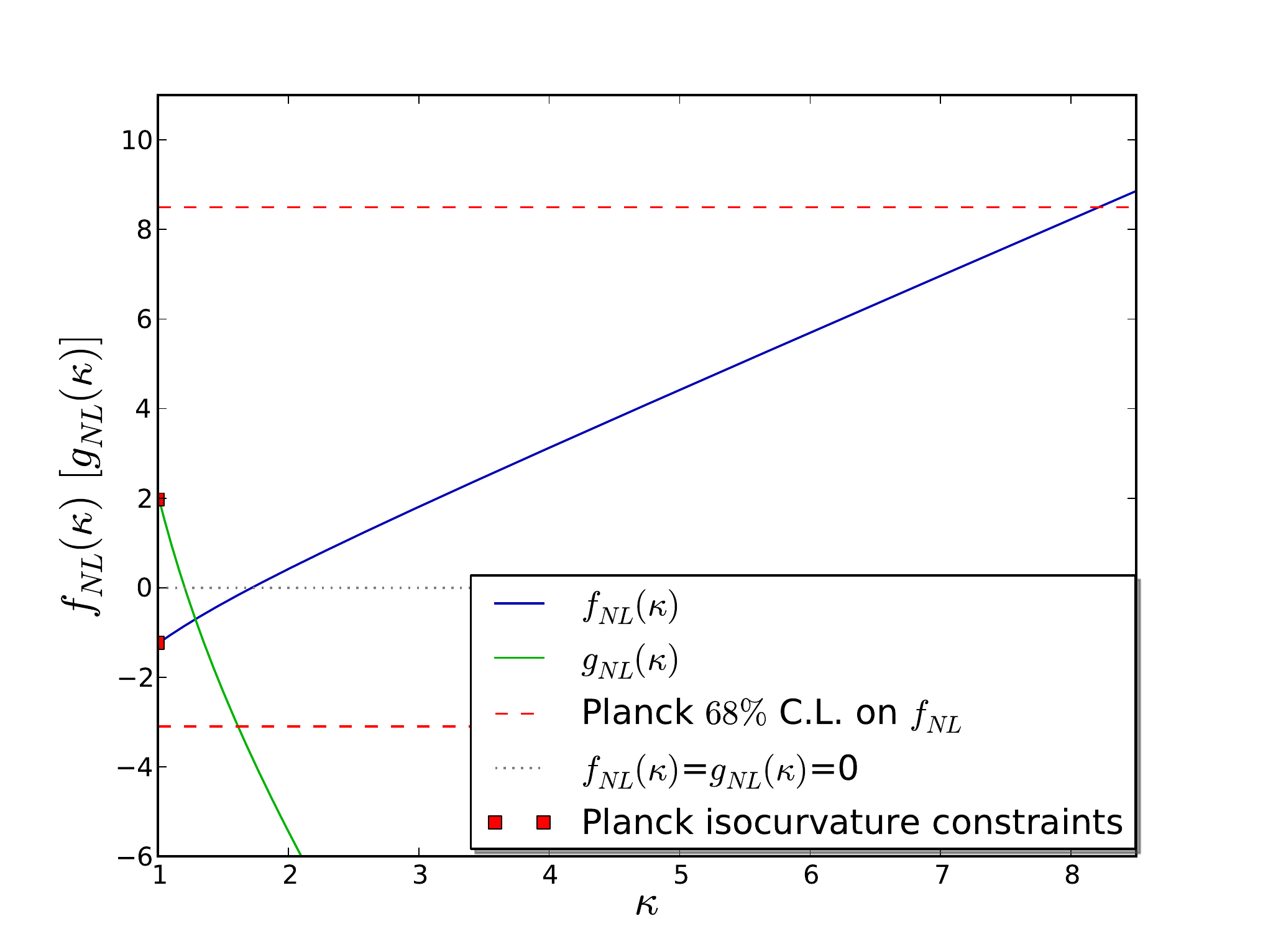}
\includegraphics[width=.7\textwidth]{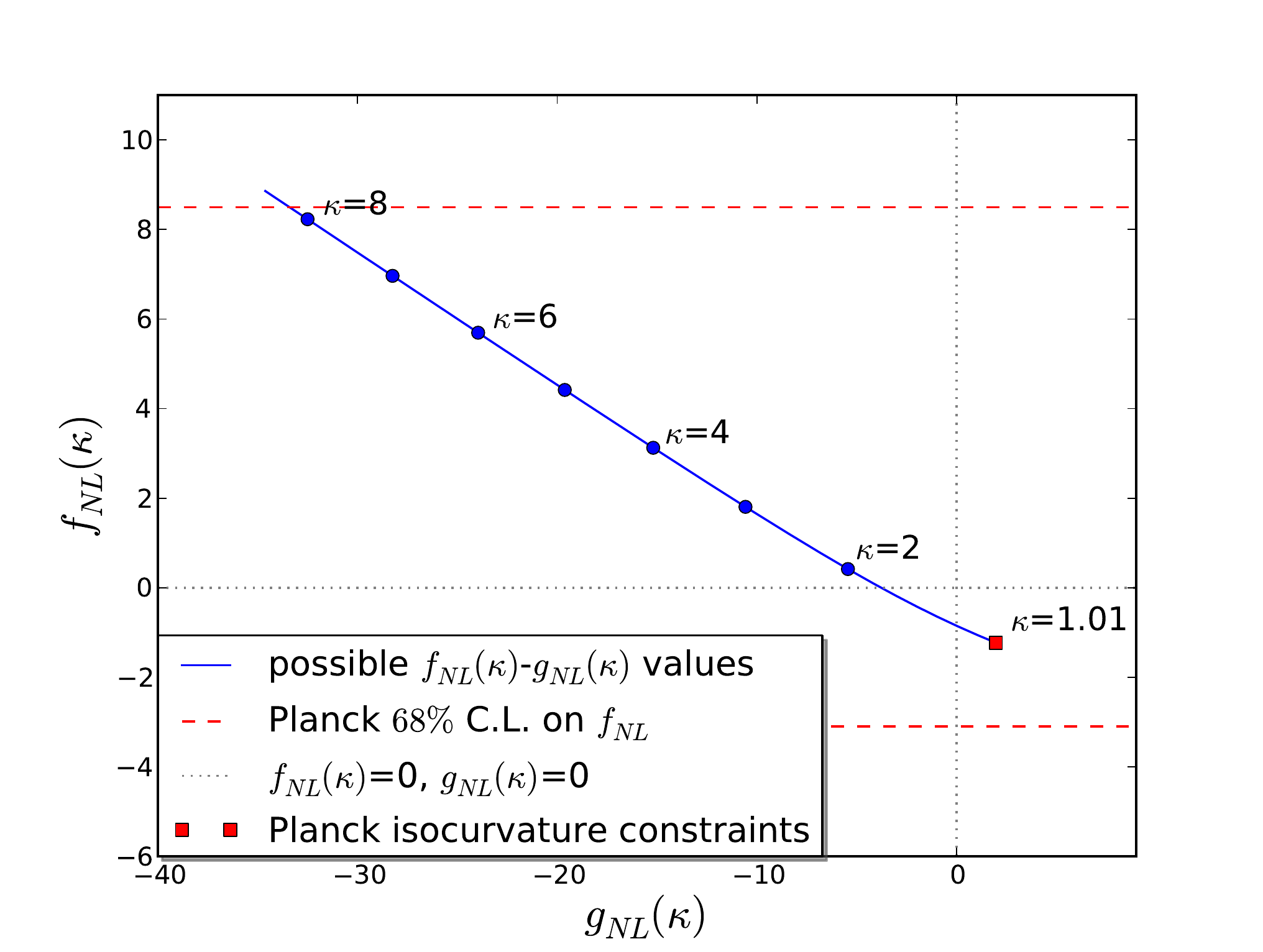}
\caption[width=\columnwidth]{Simplest curvaton model: Possible values of $f_\text{NL}$ and $g_\text{NL}$ within current \textit{Planck} constraints parametrized by the curvaton parameter $\kappa=4\bar{\rho}_r/(3\bar{\rho}_\chi) +1$. The current constraints on isocurvature modes (red squares) narrow down the allowed region significantly.\label{planck_plane}}
\end{center}
\end{figure}

Now considering the simplest model and neglecting any nonlinear evolution of $\chi$ between Hubble exit and the start of curvaton oscillations, so that $g''=0=g'''$. In this case we obtain 
\begin{equation}
\label{fg_curv}
\begin{split}
f_\text{NL}=&\frac{5}{4}\kappa-\frac{5}{3}-\frac{5}{6\kappa},\\
g_\text{NL}=&\frac{25}{54}\left(-9\kappa +\frac{1}{2} +\frac{10}{\kappa}+\frac{3}{\kappa^2}\right),
\end{split}
\end{equation}
with
\begin{equation}
\label{kapinterval}
\kappa \equiv \frac{1}{r} = \frac{4\bar{\rho}_r}{3\bar{\rho}_\chi} +1 \in [1,\infty),
\end{equation}
where the parameter $\kappa$ was introduced for reasons that become clear in Sec.\ \ref{pps}. Eq.\ (\ref{fg_curv}) is illustrated in Fig.\ \ref{planck_plane}. Currently, an upper bound on the isocurvature contribution was given by the \textit{Planck} collaboration corresponding to $f_\text{NL} = -1.23 \pm 0.02$. Note that this constraining interval for $f_\text{NL}$ was found by a power spectrum fit including adiabatic and isocurvature modes \cite{2013arXiv1303.5082P} and thus is independent of the limit ($f_\text{NL} = 2.7 \pm 5.8$) found in Ref.\ \cite{2013arXiv1303.5084P} and is only valid for the here considered curvaton scenario. On the other hand this corresponds to the interval $g_\text{NL} = 1.97 \pm 0.11$. A $g_\text{NL}$ outside this interval would put some pressure on this simplest curvaton model.~$\blacktriangleleft$
\subsection{Modulated Higgs inflation}\label{MHI}
$\blacktriangleright$ Next, we consider the Standard Model Higgs field $h$ in addition\footnote{It is well known that the Standard Model Higgs field cannot serve as an inflaton field \cite{2013PhRvD..87d3501C,2008PhRvD..77b5034I}.} to the inflaton field $\phi$ with related potential $V(\phi)$ as pointed out in Ref.\ \cite{2013PhRvD..87d3501C} and as representative mechanism of the Higgs inflation (HI) class. The Higgs field is responsible for modulating the efficiency of reheating, whereby primordial curvature perturbations are generated by converting isocurvature perturbations (produced by $h$ during inflation) to adiabatic ones \cite{2010JCAP...11..037A}. In particular we assume a simple Higgs potential during the energy scale of inflation $\mu$, 
\begin{equation}
V(h) = \frac{\lambda}{4}h^4,
\end{equation} 
with $\lambda\equiv \lambda(\mu)\approx\mathcal{O}(10^{-2})$ the Higgs self coupling with logarithmic dependence on the energy scale. 

Within this model we can write the total decay rate of the inflaton, $\Gamma(h)$, as a sum of a Higgs dependent and independent term,
\begin{equation}
\Gamma(h) = \Gamma^I + \Gamma^D(h).
\end{equation} 
Then the curvature perturbation is given by \cite{2013PhRvD..87d3501C,2008PhRvD..78b3513I}
\begin{equation}
\zeta = \frac{1}{M_\text{Pl}^2}\frac{V(\phi)}{V_\phi(\phi)}\delta\phi_* + Q_h\delta h_* + \frac{1}{2}Q_{hh}\delta h^2_* +\frac{1}{6}Q_{hhh}\delta h^3_* +\mathcal{O}(\delta h_*^4),
\end{equation} 
with $Q\approx a_0\log\left(\frac{\Gamma}{H_c}\right)$. $*$ denotes the horizon exit, $H_c$ is the Hubble constant at $t_c$ (a time before the decay of the inflaton, for details cf.\ \cite{2013PhRvD..87d3501C}), and subscript letters represent derivatives. $a_0$ is a model dependent constant of the order of $\mathcal{O}(10^{-1})$.

\begin{figure}[ht]
\begin{center}
\includegraphics[width=.7\textwidth]{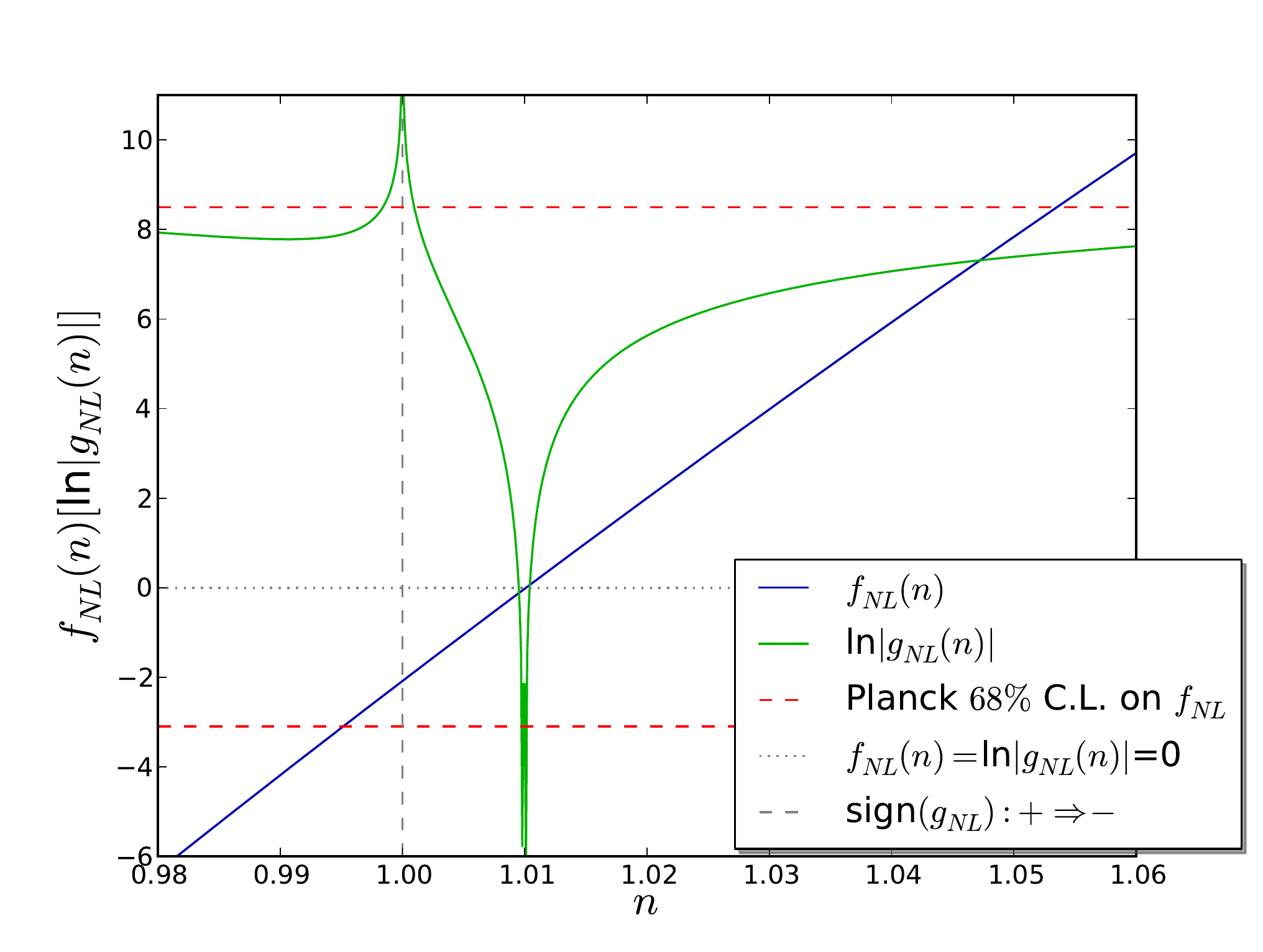}
\includegraphics[width=.7\textwidth]{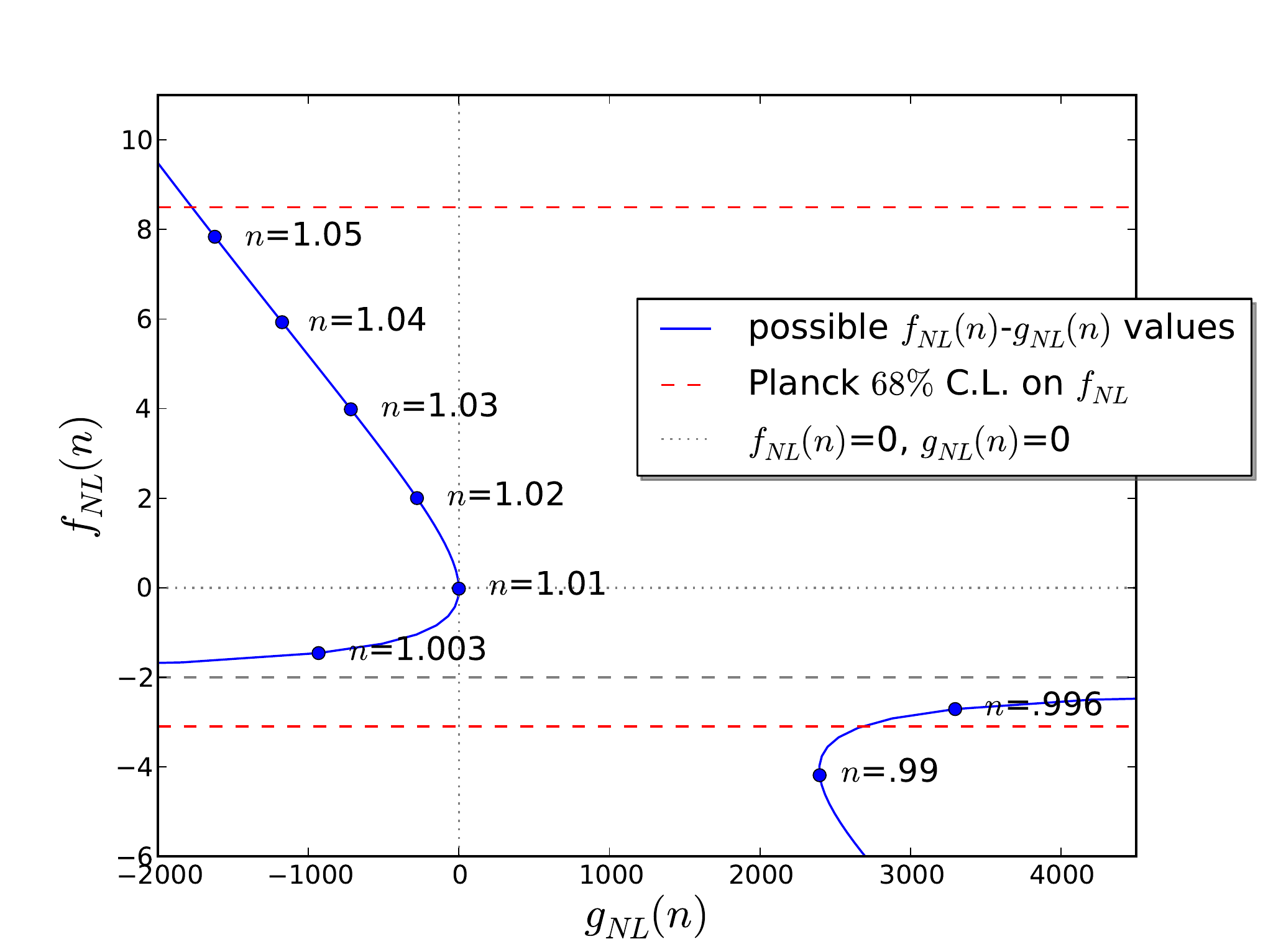}
\caption[width=\columnwidth]{Modulated Higgs inflation: Possible values of $f_\text{NL}$ and $g_\text{NL}$ for some model-typical values of $a_0=0.1,~\beta=0.5$, and $B_h = 0.01$ within the \textit{Planck} constraints parametrized by the decay rate index $n$ from $\Gamma^D(h)\propto h^n$.\label{fig:fg_hi}}
\end{center}
\end{figure}

Assuming the Higgs dependent decay rate to be of polynomial form in $h$, $\Gamma^D(h)\propto h^n$, the non-Gaussianity parameters can be calculated from the statistics of $\zeta$,  which yields \cite{2013PhRvD..87d3501C}

\begin{equation}
\label{fg_hi}
\begin{split}
f_\text{NL} = & -\frac{5}{6}\frac{\beta^2}{a_0}\left(1-\frac{\Gamma \Gamma_{hh}}{\Gamma^2_h}\right) \approx -\frac{5}{6}\frac{\beta^2}{a_0}\left(1 - \frac{1}{B_h}\frac{n-1}{n}\right),\\
g_\text{NL} = & -\frac{25 \beta^3}{54 a_0^2}\left(2 - 3\frac{\Gamma \Gamma_{hh}}{\Gamma^2_h} + \frac{\Gamma^2 \Gamma_{hhh}}{\Gamma^3_h}\right) \approx \frac{2(n-2)}{3(n-1)\beta}f_\text{NL}^2 - \frac{5}{3}\frac{\beta}{a_0}f_\text{NL},
\end{split}
\end{equation}

\noindent with $B_h = \Gamma^D/\Gamma \leq \mathcal{O}(10^{-3} - 10^{-2})$ and $\beta \approx \mathcal{O}(10^{-2} - 1)$. Eq.\ (\ref{fg_hi}) is illustrated in~Fig.~\ref{fig:fg_hi}.~$\blacktriangleleft$

\begin{commenta}
\subsection{Small Field Inflation (SFI)}
...still to discuss

\subsection{Overview}
The dependence of $f_\text{NL}$ and $g_\text{NL}$ on the previously discussed inflation models is summarized in Tab.\ \ref{table1}.
\begin{table}[ht]
\begin{center}
\begin{tabular}{|c|c|c|c|}
\hline
\text{Inflation model}& $p$ & $f_\text{NL}(p)$ & $g_\text{NL}(p)$\\
\hline
CI	& $\kappa$ & $\frac{5}{4}\kappa-\frac{5}{3}-\frac{5}{6\kappa}$ & $\frac{25}{54}\left(-9\kappa +\frac{1}{2} +\frac{10}{\kappa}+\frac{3}{\kappa^2}\right)$\\			
\hline
HI & $n$ & $-\frac{5}{6}\frac{\beta^2}{a_0}\left(1 - \frac{1}{B_h}\frac{n-1}{n}\right)$ & $ \frac{2(n-2)}{3(n-1)\beta}f_\text{NL}^2 - \frac{5}{3}\frac{\beta}{a_0}f_\text{NL}$\\
\hline
SFI &&&\\
\hline
\end{tabular}
\caption{\label{table1}Summary of the discussed inflation scenarios. The table shows the particular inflation model and its corresponding, model parametrized expressions for $f_\text{NL}(p)$ and $g_\text{NL}(p)$.}
\end{center}
\end{table}

\end{commenta}
\section{Posterior for special inflationary parameters}\label{sec:validation}
\subsection{Generic procedure} \label{subsec:gen}
For all models discussed in Sec.\ \ref{sec:special_models} an expression for the posterior of a model specific quantity can be derived by replacing $f_\text{NL}$ and $g_\text{NL}$ by their corresponding model dependent parameters $p$, which are pointed out in the stated section. This is, of course, also true for all other inflation models postulating these two non-Gaussianity parameters. Thereby one obtains $p$-dependent equations for $\frac{\delta H(\zeta_1,d|p)}{\delta \zeta_1}\bigg\vert_{\zeta_1=\bar{\zeta_1}}=0$ and $D^{-1}_{d,p}\equiv \frac{\delta^2 H(\zeta_1,d|p)}{\delta \zeta_1^2}\bigg\vert_{\zeta_1=\bar{\zeta_1}}$, which allow to derive the posterior analytically, Eq.\ (\ref{postgen}). 

Eventually, the response $R$ has to be replaced by its respective, corresponding expression, depending on whether one uses the CMB data, LSS data, or something else. The resulting posterior can then be implemented and evaluated numerically. With its help one might obtain a single point (e.g.\ mean of $p$ or maximum of the posterior) within the $f_\text{NL}(p)-g_\text{NL}(p)-$plane with corresponding error interval being a sub-area of the plane or it just maps out the parameter posterior.

\subsection{Simplest curvaton model}\label{sec:scmexp}
\subsubsection{Posterior derivation}
To demonstrate the applicability of the inference approach we study the simplest curvaton inflation scenario. We follow Secs.\ \ref{sec:ift} and \ref{subsec:gen} to derive the posterior. The replacement of $f_\text{NL}$ and $g_\text{NL}$ by their corresponding $\kappa$ expressions leads to the information Hamiltonian,

\begin{equation}
\label{hamkap}
\begin{split}
H&(\zeta_1,d|\kappa) = H_0 + \frac{1}{2}\zeta^\dag_1 D^{-1}\zeta_1 - j^\dag\zeta_1 -\left(\frac{3}{4}\kappa -1 -\frac{1}{2\kappa}\right) j^\dag \zeta_1^2 - \left(-\frac{3}{2}\kappa + \frac{1}{12} +\frac{5}{3\kappa} +\frac{1}{2\kappa^2}\right) j^\dag \zeta^3_1\\
	& +\left(\frac{3}{4}\kappa -1 -\frac{1}{2\kappa}\right) \zeta_1^\dag M \zeta_1^2 +\left(-\frac{3}{2}\kappa + \frac{1}{12} +\frac{5}{3\kappa} +\frac{1}{2\kappa^2}\right) \zeta^\dag_1 M \zeta^3_1\\
	& + \frac{1}{2} \left(\frac{3}{4}\kappa -1 -\frac{1}{2\kappa}\right)^2 \left(\zeta^2_1\right)^\dag M \zeta_1^2  +\left(-\frac{3}{2}\kappa + \frac{1}{12} +\frac{5}{3\kappa} +\frac{1}{2\kappa^2}\right)\left(\frac{3}{4}\kappa -1 -\frac{1}{2\kappa}\right) \left(\zeta^2_1\right)^\dag M \zeta_1^3\\
	& + \frac{1}{2}\left(-\frac{3}{2}\kappa + \frac{1}{12} +\frac{5}{3\kappa} +\frac{1}{2\kappa^2}\right)^2 \left(\zeta^3_1\right)^\dag M \zeta_1^3 .
\end{split}
\end{equation}
Analogously, by replacing $f_\text{NL}$ and $g_\text{NL}$ by their corresponding $\kappa$ expressions in Eqs.\ (\ref{gradgen}) and (\ref{hessgen}), one obtains expressions for $\frac{\delta H(\zeta_1,d|\kappa)}{\delta \zeta_1}\bigg\vert_{\zeta_1=\bar{\zeta_1}}=0$ and $D^{-1}_{d,\kappa}\equiv \frac{\delta^2 H(\zeta_1,d|\kappa)}{\delta \zeta_1^2}\bigg\vert_{\zeta_1=\bar{\zeta_1}}$, whereby we are able to perform the saddle-point approximation of the posterior, which yields (see Eq.\ (\ref{postgen}))

\begin{commenta}
\textbf{Leading-order $\kappa$.} For the information Hamilton $H(\zeta_1,d|\kappa)$ in leading-order $\kappa$ approximation we obtain

\begin{equation}
\label{ham}
\begin{split}
H(\zeta_1,d|\kappa) =&~ H_0 + \frac{1}{2}\zeta^\dag_1 D^{-1}\zeta_1 - j^\dag\zeta_1 - \frac{3}{4}\kappa j^\dag \zeta_1^2 +\frac{3}{2}\kappa j^\dag \zeta^3_1 +\frac{3}{4}\kappa \zeta_1^\dag M \zeta_1^2 \\
 		   &-\frac{3}{2}\kappa \zeta^\dag_1 M \zeta^3_1 + \frac{9}{32}\kappa^2 \left(\zeta^2_1\right)^\dag M \zeta_1^2  -\frac{9}{8}\kappa^2 \left(\zeta^2_1\right)^\dag M \zeta_1^3 + \frac{9}{8}\kappa^2 \left(\zeta^3_1\right)^\dag M \zeta_1^3 ,
\end{split}
\end{equation}

\noindent with related gradient

\begin{equation}
\label{grad}
\begin{split}
0=&~\frac{\delta H(\zeta_1,d|\kappa)}{\delta \zeta_1}\bigg\vert_{\zeta_1=\bar{\zeta_1}} = \left(D^{-1}-\frac{3}{2}\kappa ~\hat{j}\right)\bar{\zeta_1} - j + \frac{9}{2}\kappa \hat{j}\bar{\zeta_1}^2 +\left(\frac{3}{4}\kappa M \bar{\zeta_1}^2 +\frac{3}{2} \kappa\bar{\zeta_1} \star M\bar{\zeta_1}\right)  \\
 & + \left(-\frac{3}{2}\kappa M \bar{\zeta_1}^3 - \frac{9}{2}\kappa \bar{\zeta_1}^2\star M\bar{\zeta_1}\right) + \left(\frac{9}{8}\kappa^2 \bar{\zeta_1}\star M \bar{\zeta_1}^2\right)\\
 & + \left(-\frac{9}{4}\kappa^2\bar{\zeta_1} \star M \bar{\zeta_1}^3 - \frac{27}{8}\kappa^2\bar{\zeta_1}^2 \star M\bar{\zeta_1}^2\right) + \left(\frac{27}{4}\kappa^2 \bar{\zeta_1}^2 \star M\bar{\zeta_1}^3\right), 
\end{split}
\end{equation}

\noindent and Hessian

\begin{equation}
\label{hess}
\begin{split}
D^{-1}_{d,\kappa}&\equiv \frac{\delta^2 H(\zeta_1,d|\kappa)}{\delta \zeta_1^2}\bigg\vert_{\zeta_1=\bar{\zeta_1}} = 
  D^{-1} -\frac{3}{2}\kappa\hat{j} + 9\kappa \bar{\zeta_1}\star \hat{j} +\frac{3}{2}\kappa\left(2\bar{\zeta_1}\star M + \widehat{M\bar{\zeta_1}}\right)\\
	 &~-\frac{9}{2}\kappa\left(\widehat{M\bar{\zeta_1}^2} +\left(\bar{\zeta_1}^2\star M + 2\widehat{{\bar{\zeta_1}\star M\bar{\zeta_1}}}\right)\right) +\frac{9}{8}\kappa^2\left(\widehat{M\bar{\zeta_1}^2}+2\bar{\zeta_1}\star M\star \bar{\zeta_1}\right)\\
	&~-\frac{9}{4}\kappa^2\left(3\widehat{\bar{\zeta_1}\star M\bar{\zeta_1}^2} +\widehat{M\bar{\zeta_1}^3} +6\bar{\zeta_1}\star M\star \bar{\zeta_1}^2\right) +\frac{27}{2}\kappa^2\left(\widehat{\bar{\zeta_1}\star M\bar{\zeta_1}^3} +\frac{3}{2}\bar{\zeta_1}^2\star M\star \bar{\zeta_1}^2\right).
\end{split}
\end{equation}

\noindent whereby we are able to perform the saddle-point approximation of the posterior, which yields (see Eq.\ (\ref{postgen}))
\end{commenta}
\begin{equation}
\begin{split}
P(\kappa|d)\propto&~ P(\kappa)\int \mathcal{D}\zeta_1 \exp\left[-H(d,\zeta_1|\kappa)\right] \approx |2\pi D_{d,\kappa}|^{\frac{1}{2}}\exp\left[-H(d,\bar{\zeta_1}|\kappa)\right]P(\kappa).
\end{split}
\end{equation}
For numerical reasons $D^{-1}_{d,\kappa}$ is split into a diagonal part, $D^{-1}_{d,\kappa,\text{diag}}$, and a non-diagonal one, $D^{-1}_{d,\kappa,\text{non-diag}}$, which leads to (cf.\ \cite{paper1})

\begin{equation}
\label{post}
\begin{split}
&\ln\left[P(\kappa|d)\right]=-H(\kappa|d)
\approx -\frac{1}{2}\text{tr}\left[\ln\left(\frac{1}{2\pi} D_{d,\kappa,\text{diag}}^{-1}\right) \right]\\
&~~~~ +\frac{1}{2}\text{tr}\left[ \sum_{n=1}^\infty \frac{(-1)^n}{n} \left(D_{d,\kappa,\text{diag}}D_{d,\kappa,\text{non-diag}}^{-1}\right)^n \right] -H(d,\bar{\zeta_1}|\kappa) +\ln\left[P(\kappa)\right]+\text{const.}.
\end{split}
\end{equation}
The series expansion of the logarithm in Eq.\ (\ref{post}) can be truncated if the terms become sufficiently small.

\subsubsection{Numerical implementation}\label{sec:impl}
For a numerical implementation of Eq.\ (\ref{post}) we have to choose a prior pdf for $\kappa$, $P(\kappa)$. A naive choice, for instance, according to Eq.\ (\ref{kapinterval}) would be \begin{equation}
\label{prior}
P(\kappa) = \frac{1}{\kappa_0 -1}\Theta(\kappa -1)\Theta(\kappa_0-\kappa),
\end{equation}
with $\Theta$ the Heaviside step function and $\kappa_0$ a large but finite number to normalize the distribution. However, at this point we do not want to open a discussion of how to choose an appropriate prior pdf for the curvaton parameter. Thus we focus on the likelihood pdf, $P(d|\kappa)\propto \frac{P(\kappa|d)}{P(\kappa)}$, given by subtracting $\ln\left[P(\kappa)\right]$ from Eq.\ (\ref{post}).

In order to implement the likelihood we study the inference from CMB data in the Sachs-Wolfe limit \cite{1967ApJ...147...73S} within a toy example in two and three dimensions. Here the data are given by
\begin{equation}
\begin{split}
d =&~ R\zeta +n = R_\text{CMB}\Phi_\text{md} +n = R_\text{CMB}\frac{3}{5}\zeta +n\\
  =&~ R_\text{CMB}\frac{3}{5}\left(\zeta_1 + \frac{3}{5}f_\text{NL}\zeta_1^2 +\frac{9}{25}g_\text{NL}\zeta_1^3 +\mathcal{O}(\zeta_1^4)\right) +n,
\end{split}
\end{equation}
with $\Phi_\text{md}$ the Bardeen potential in the matter dominated era. In this limit the response becomes local\footnote{The treatment of non-local responses was shown in \cite{paper1} for a similar case.} \cite{1997A&A...321....8W},
\begin{equation}
R(x,y) = -\frac{3}{5}\frac{1}{3}\delta(x-y),
\end{equation}
with $x,y$ two positions on the two-(three-) dimensional sky. Additionally, we assume white noise, $N_{xy}=\sigma_n^2 \delta_{xy}$. For the inference process the so-called free information propagator $D$ is required, which depends in particular on the power spectrum of $\zeta_1$. For this we assume
\begin{equation}
\label{pex}
\Xi = \left\langle \zeta_1\zeta_1^\dag\right\rangle_{(\zeta_1|\Xi)} \equiv P_{\zeta_1}(k)\delta_{kk'} = A_s \left(\frac{k}{k_*}\right)^{n_s -1}\delta_{kk'},
\end{equation}
which is diagonal in Fourier space with related modes $k,k'$. This power spectrum is parametrized by the scalar amplitude $A_s$, the spectral index $n_s$, and the pivot scale $k_*$. A detailed discussion about this power spectrum and its parameters can be found in Sec.\ \ref{sec:pps} and Ref.\ \cite{2013arXiv1303.5082P}. 

The numerical implementation is done in \textsc{NIFTy} \cite{2013arXiv1301.4499S}, where possible calculations of traces of operators are determined by operator probing. The \textsc{NIFTy} package uses implicit operators and therefore avoids to store matrices explicitely. Thus, highly resolved data sets like the CMB map of the \textit{Planck} satellite should be treatable in principle. However, to show the efficiency of the derived inference method we use a data set in two (three) flat dimensions with best fit parameters (for scalar amplitude and spectral index from \textit{Planck} \cite{2013arXiv1303.5082P}) $A_s = 2.2\times10^{-9}$, $n_s = 0.9603$, $k_*=1$, $N_\text{pix}=10000$ $(10648)$, and $\sigma_n^2 = 10^{-14}$, where $\zeta_1$ and $n$ are sampled from $\Xi$ and $N$, respectively.
\begin{figure}[ht]
\begin{center}
\includegraphics[width=.5\textwidth]{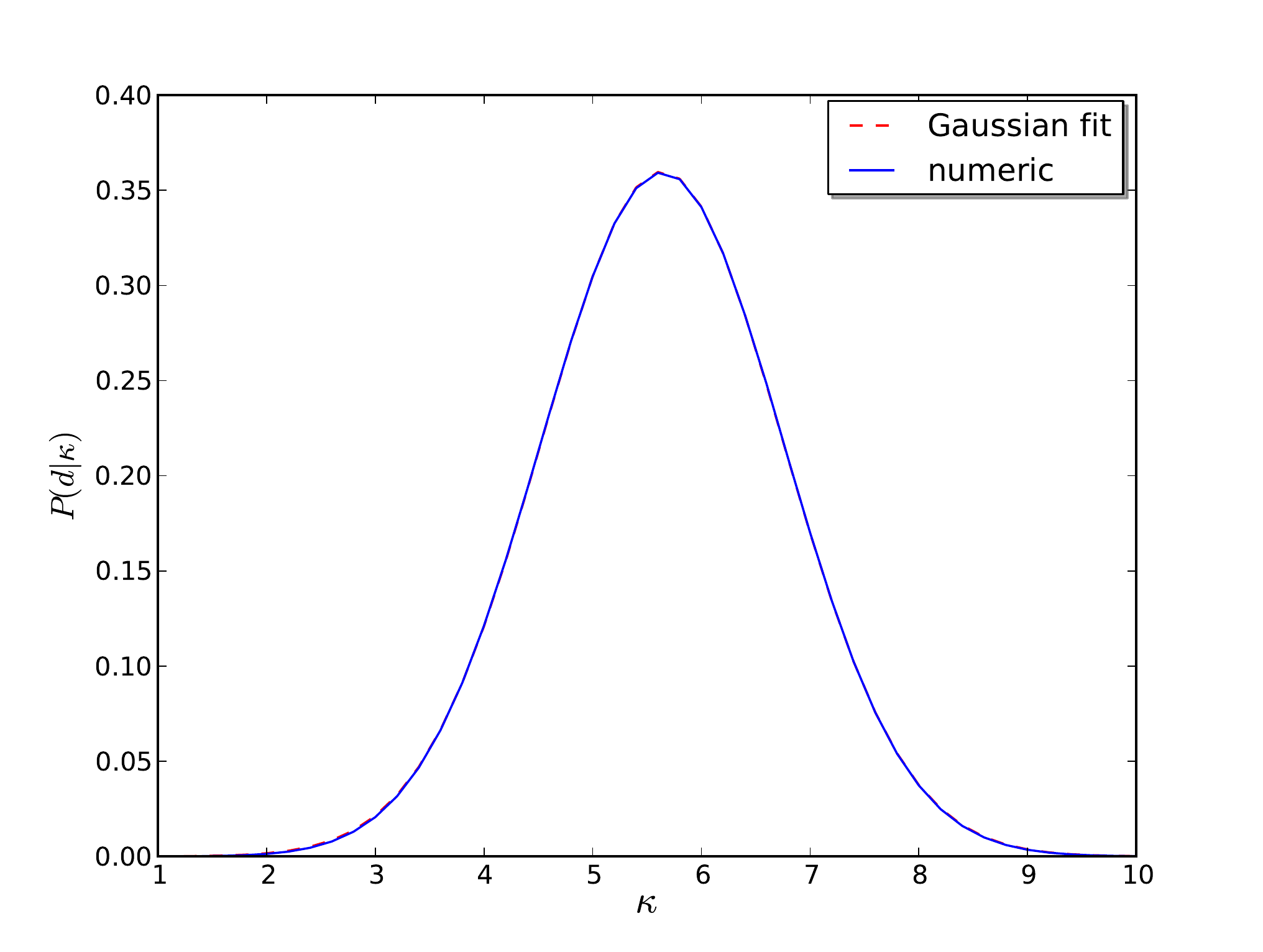}%
\includegraphics[width=.5\textwidth]{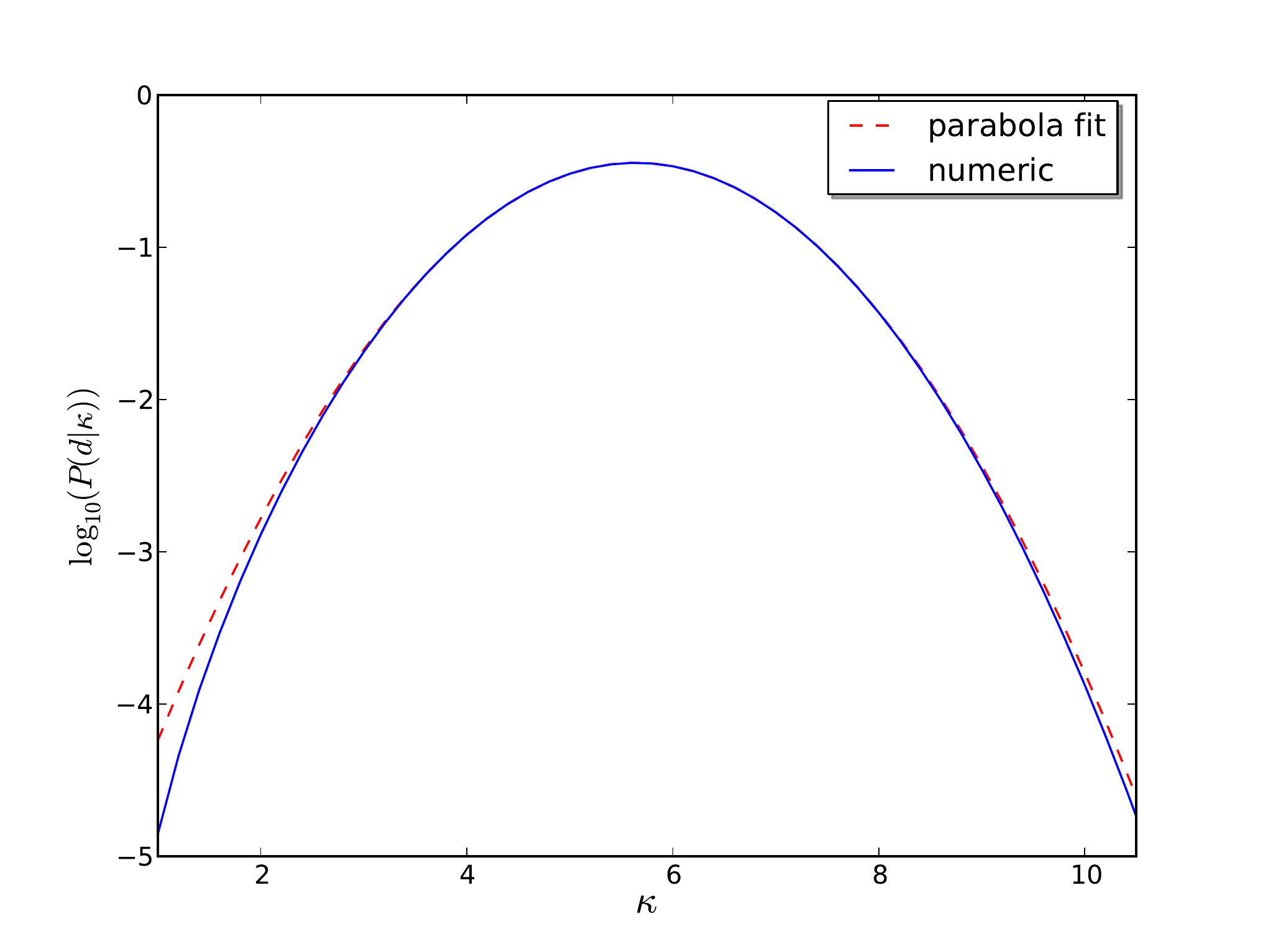}\\
(a)\hspace{7.5cm}(b)\\
\includegraphics[width=.5\textwidth]{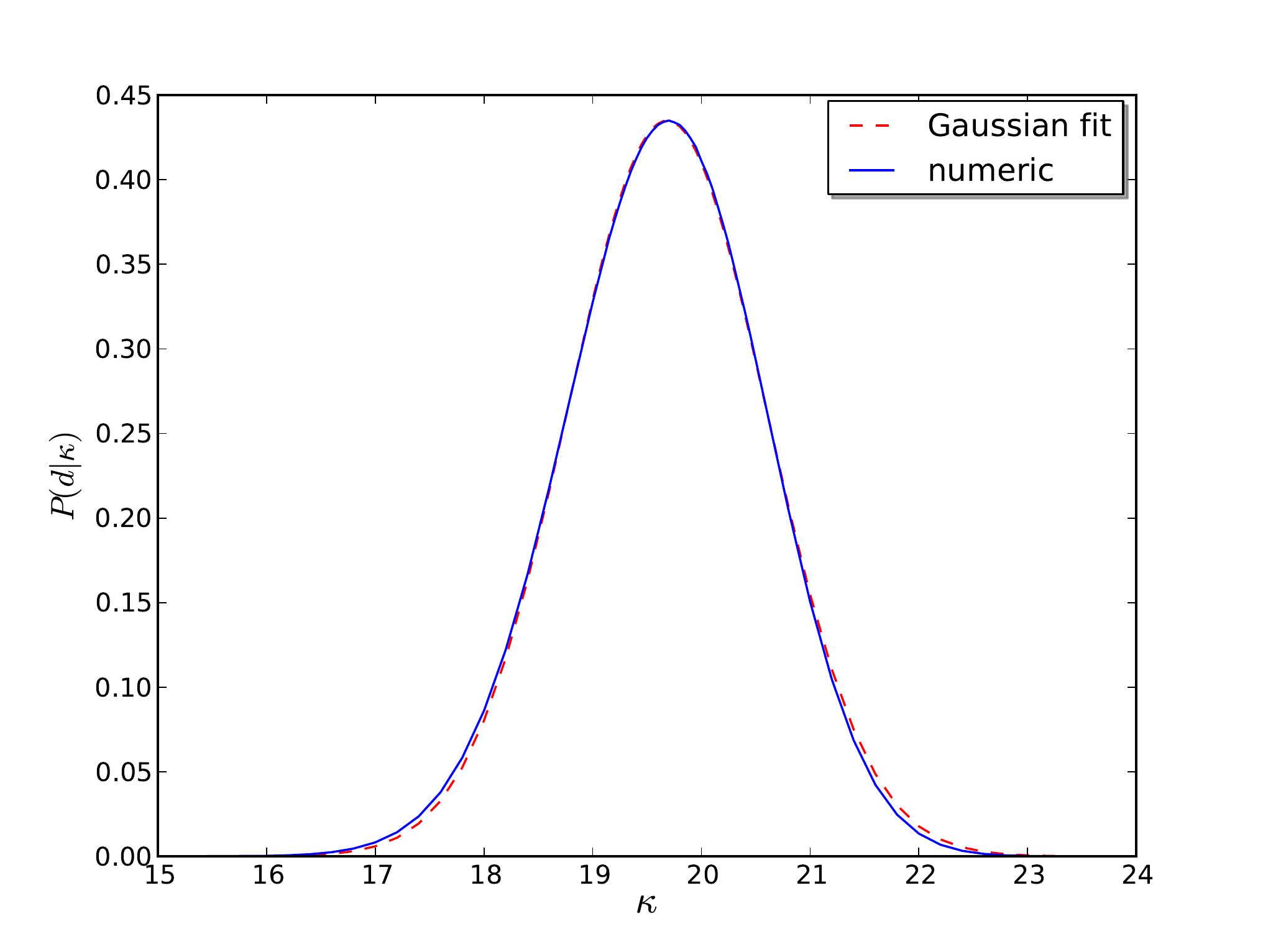}%
\includegraphics[width=.5\textwidth]{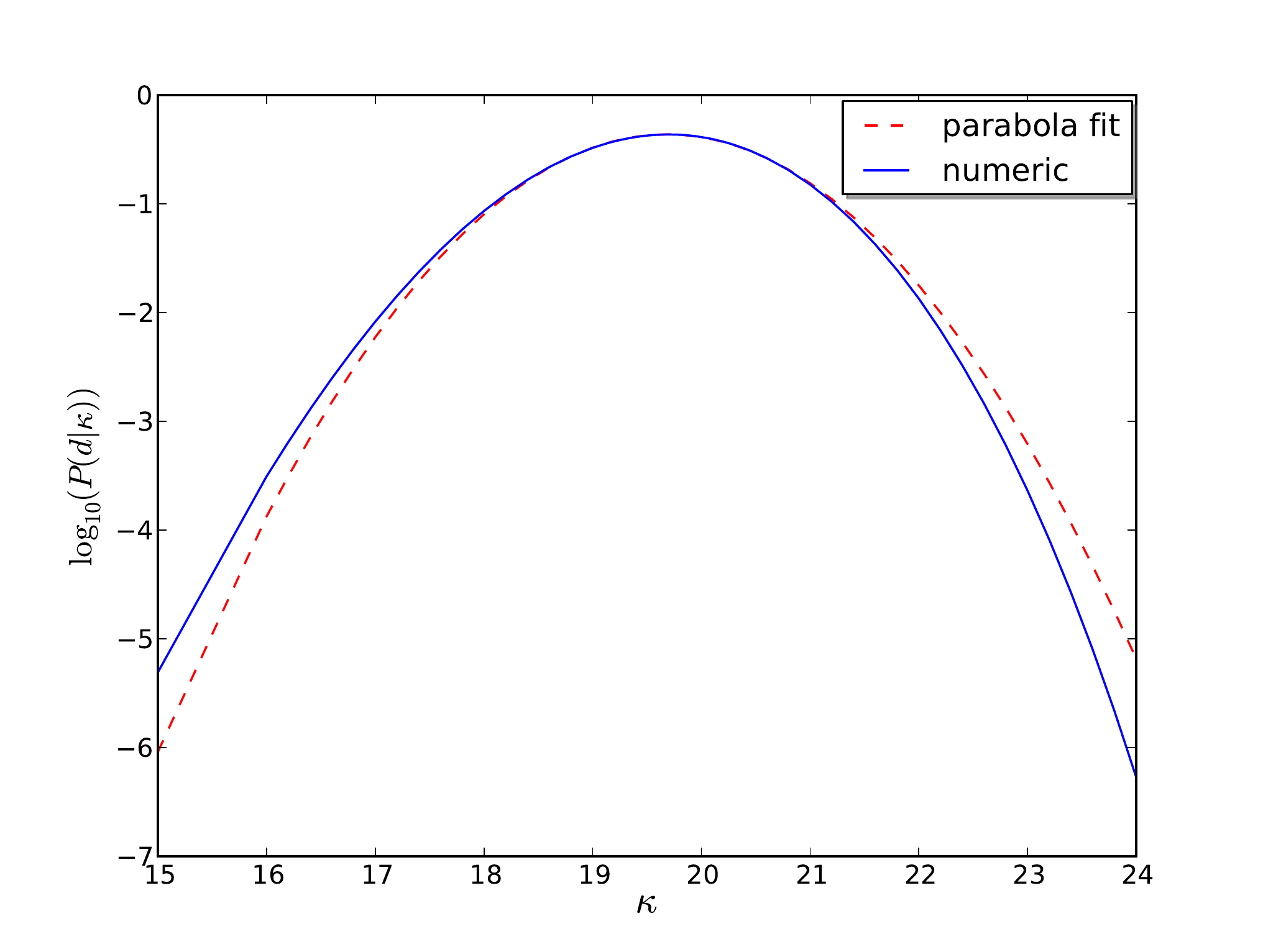}\\
(c)\hspace{7.5cm}(d)
\end{center}
\caption[width=\columnwidth]{(Color online) Normalized likelihood distributions for $\kappa$ in a two-(three-)dimensional test case with data generated from $\kappa_\text{gen}=5$ [(a), (b)] ($\kappa_\text{gen}=19.8$ [(c), (d)]).}
\label{shape}
\end{figure}

An implementation of the likelihood (posterior\footnote{For a prior choice according to Eq.\ (\ref{prior}), Fig.\ \ref{shape} shows the posterior of $\kappa$.} with constant prior) in two (three) dimensions with a true underlying value of $\kappa_{\text{gen}}=5$ ($\kappa_\text{gen}=19.8$) is shown in Fig.\ \ref{shape}. The numerical result coincides [perfectly, (a)] with a Gaussian fit and would deviate from this shape only for unrealistically high values of $\kappa$. A slight deviation from Gaussianity, however, can even be observed in the three-dimensional realization with $\kappa=19.8$ (likelihood is negatively skewed) \cite{2007JCAP...03..019C,2011PhRvD..84f3013S,2010ApJ...724.1262E,2010A&A...513A..59E,2009PhRvD..80j5005E,PhysRevD.87.063003,KSW}. As we will show in Sec.\ \ref{sec:toy}, even this slight deviation (and even smaller ones) can affect the reconstruction of the primordial power spectrum, $P_\zeta(k)$, significantly. Note, however, that this statement is only true for the likelihood of $\kappa$ and has to be taken into account case by case. Significant non-Gaussianity in the likelihood and posterior pdf might be induced by the $p$ dependent determinant of Eq.\ (\ref{postgen}) or by a specific prior choice. A brief discussion about the comparison of a skewed posterior of $p$ can be found in App. \ref{app}. 

\subsubsection{Posterior validation}\label{sec:valid}
To validate the implementation of Eq.\ (\ref{post}) we consider the two-dimensional test case of Sec.\ \ref{sec:impl} and apply the DIP test \cite{paper2} by following Refs.\ \cite{paper1,Cook06validationof}, i.e.\ conducting the following steps:

\begin{enumerate}
	\item Sample uniformly a value of $\kappa_{\text{gen}}$ from an interval $I=\left[\kappa_\text{ini},\kappa_\text{fin}\right]$, i.e. from a prior\footnote{Note that for validating both, the sufficiency of numerical implementation of the posterior distribution and analytic approximations including its derivation, it is not necessary to choose a physical prior. Thus this kind of prior with appropriate values $\kappa_\text{ini, fin}$ has been chosen for simplicity only. Here, appropriate means that the interval $I$ is sufficiently large to take care of the shape of the posterior.} \\
    		\begin{equation}
		P(\kappa)=\left\{
    		\begin{array}{cc}
                		 \frac{1}{\kappa_\text{fin}-\kappa_\text{ini}} &~~~~~~~\text{if}~ \kappa\in I\\
                 		 0 		& ~\text{else}
    		\end{array} 
    		\right..
		\end{equation}
	\item Generate data $d$ for $\kappa_{\text{gen}}$ according to Eq.~(\ref{data}).
	\item Calculate a posterior curve for given data by determining $P(\kappa|d)$ for $\kappa \in I$ according to Eq. (\ref{post}).
	\item Calculate the posterior probability for $\kappa\leq \kappa_\text{gen}$ according to 
	
\begin{equation}
\label{check}
x\equiv \int_{\kappa_\text{ini}}^{\kappa_{\text{gen}}} \text{d}\kappa~ P(\kappa|d) ~\in \left[0,1\right].
\end{equation}  

	\item If the calculation of the posterior was correct, the distribution for $x$, $P(x)$, should be uniform between 0 and 1.
\end{enumerate}

\begin{figure}[ht]
\begin{center}
\includegraphics[width=.6\textwidth]{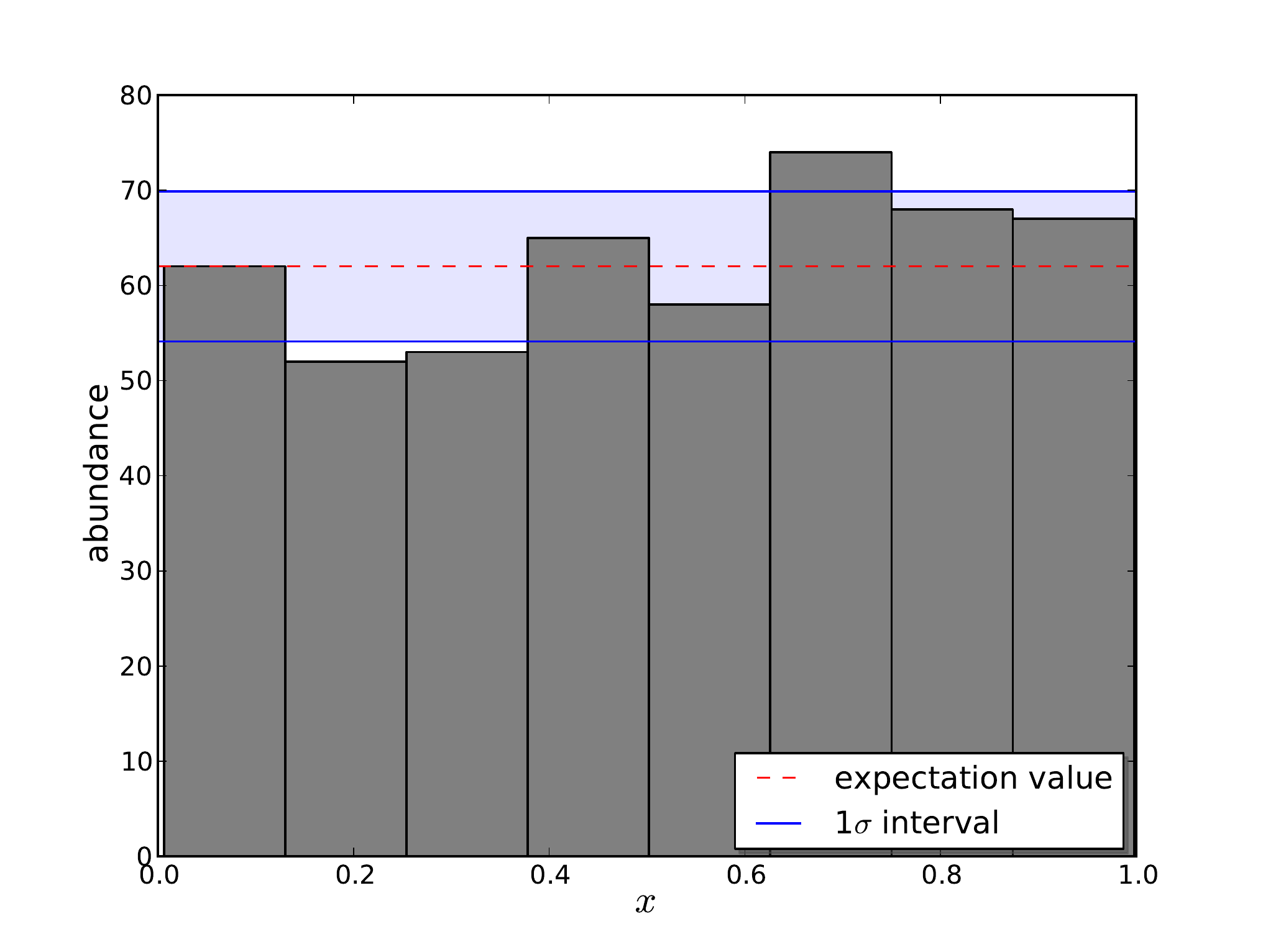}%
\end{center}
\caption[width=\columnwidth]{(Color online) DIP distributions of calculated $x$ values for the two-dimensional test case. The histogram shows the unnormalized distribution of 500 $x$ values within eight bins. The standard deviation interval ($1\sigma$, blue solid line) around the expectation value (red dashed line) as calculated from Poissonian statistics is also shown.}
\label{dip1}
\end{figure}

The result of this posterior validation test is shown by Fig.\ \ref{dip1}. Here, the histogram represents the distribution of 500 $x$-values within eight bins. The uniformity of the distribution verifies the numeric and analytic (due to the saddle-point approximation) sufficiency of the posterior. In particular, this means that the shape of the posterior (and therefore the error-bars around the posterior mean) are calculated correctly. Otherwise the histogram of the DIP test would have exhibited a characteristic deviation from uniformity, e.g., a dip in the case of an underestimation of the variance.

\begin{commenta}
\textbf{Leading $\kappa$.} A one-dimensional implementation of Eq.\ (\ref{post}) with $N_\text{pix}=1000$, $N=10^{-14}$, $R_\text{CMB}=-\frac{1}{3}$ and mock data with the CMB angular power spectrum, $C_l^\text{CMB}$, and true underlying value of $\kappa_\text{gen}=5$ ($\kappa_\text{gen}=4000$) is shown by the two upper (lower) panels of Fig.\ \ref{shape}.

\clearpage
\begin{figure}[ht]
\begin{center}
\includegraphics[width=.5\textwidth]{post_kap_5.png}%
\includegraphics[width=.5\textwidth]{post_kap_5_log.png}\\
\includegraphics[width=.5\textwidth]{post_kap_4000.png}%
\includegraphics[width=.5\textwidth]{post_kap_4000_log.png}\\
\end{center}
\caption[width=\columnwidth]{Normalized posterior distributions for $\kappa$ in a one-dimensional test case with data generated from $\kappa_\text{gen}=5$.}
\label{shape}
\end{figure}

To validate the posterior we use the DIP test, pointed out in Ref.\ \cite{paper2}. The result, which validates the sufficiency of the leading $\kappa$ approximation is shown in Fig.\ \ref{valid_1d_lead}.
\begin{figure}[ht]
\begin{center}
\includegraphics[width=.5\textwidth]{hist_500_pm1000_Ne-14.pdf}%
\caption[width=\columnwidth]{DIP test of the leading order $\kappa$ posterior in one dimension.}
\end{center}
\label{valid_1d_lead}
\end{figure}
\end{commenta}
\subsection{Modulated Higgs inflation}\label{sec:HI}
Next, we consider the scenario of modulated Higgs inflation discussed in Sec.\ \ref{MHI}. By following again Secs.\ \ref{sec:ift} and \ref{subsec:gen}, i.e.\ by replacing $f_\text{NL}$ and $g_\text{NL}$ by their corresponding $n$ dependent expressions\footnote{As a reminder: the model is parametrized by the decay rate index $n$, $\Gamma^D(h)\propto h^n$ (see Sec.\ \ref{MHI} for details).}, we obtain
\begin{equation}
\label{hamgen_n}
\begin{split}
H&(\zeta_1,d|f_\text{NL}(n))\\
 =&~ -\ln\left[ P(d|\zeta_1,f_\text{NL}(n))P(\zeta_1|f_\text{NL}(n))\right] = -\ln\left[\mathcal{G}\left(d-R\zeta,N\right)\mathcal{G}(\zeta_1,\Xi)\right]\\
=&~H_0 + \frac{1}{2}\zeta^\dag_1 D^{-1}\zeta_1 - j^\dag\zeta_1 -\frac{3}{5}f_\text{NL}j^\dag\zeta_1^2 -\left(\frac{6(n-2)}{25(n-1)\beta}f_\text{NL}^2 - \frac{3}{5}\frac{\beta}{a_0}f_\text{NL}\right)j^\dag\zeta_1^3 +\frac{3}{5}f_\text{NL}\zeta_1^\dag M \zeta_1^2\\ 
	&+\left(\frac{6(n-2)}{25(n-1)\beta}f_\text{NL}^2 - \frac{3}{5}\frac{\beta}{a_0}f_\text{NL}\right)\zeta_1^\dag M \zeta_1^3 +\frac{9}{50}f^2_\text{NL}\left(\zeta_1^\dag\right)^2 M \zeta_1^2\\
	& +\left(\frac{18(n-2)}{125(n-1)\beta}f_\text{NL}^3 - \frac{81}{625}\frac{\beta}{a_0}f^2_\text{NL}\right)\left(\zeta_1^\dag\right)^2 M \zeta_1^3 \\ 
	& +\left(\frac{27(n-2)}{625(n-1)\beta}f_\text{NL}^2 - \frac{27}{250}\frac{\beta}{a_0}f_\text{NL}\right)^2\left(\zeta_1^\dag\right)^3 M \zeta_1^3,
\end{split}
\end{equation}
with $f_\text{NL}(n)$ given by 
\begin{equation}
f_\text{NL}(n) \approx -\frac{5}{6}\frac{\beta^2}{a_0}\left(1 - \frac{1}{B_h}\frac{n-1}{n}\right).
\end{equation}
In Eq.\ (\ref{hamgen_n}) we dropped the explicit dependency of $f_\text{NL}$ on $n$ in our notation for reasons of clarity. To derive the posterior pdf for the $n$ parameter of the Higgs inflation model, we again conduct the saddle-point approximation introduced in Sec.\ \ref{sec:ift}. This yields
\begin{equation}
\label{post_n}
\begin{split}
&\ln[P(n|d)]=-H(n|d)
\approx -\frac{1}{2}\text{tr}\left[\ln\left(\frac{1}{2\pi} D_{d,n,\text{diag}}^{-1}\right) \right]\\
&~~~~ +\frac{1}{2}\text{tr}\left[ \sum_{m=1}^\infty \frac{(-1)^m}{m} \left(D_{d,n,\text{diag}}D_{d,n,\text{non-diag}}^{-1}\right)^m \right] -H(d,\bar{\zeta_1}|n) +\ln[P(n)]+\text{const.},
\end{split}
\end{equation}
where $\bar{\zeta}_1$ and $D_{d,n}$ are defined in Eqs.\ (\ref{gradgen}) and (\ref{hessgen}) and the labels \textit{diag} and \textit{non-diag} refer to the diagonal and non-diagonal part of $D^{-1}_{d,n}$. As before, the series expansion of the logarithm can be truncated if the terms become sufficiently small.

The numerical implementation of Eq.~(\ref{post_n}) and its validation is completely analogous to the one of the curvaton scenario. Therefore we do not present it here.

\section{Primordial power spectrum reconstruction}\label{sec:pps}
\subsection{Motivation}
$\blacktriangleright$ An essential quantity that allows to discriminate between inflationary scenarios is the primordial power spectrum. It is commonly parametrized by \cite{2013arXiv1303.5082P}
\begin{equation}
\label{powerdef}
\begin{split}
\ln\left[ P_\mathcal{R}(k)\right] =&~ \ln( A_s) + \left(n_s -1 +\frac{1}{2}\frac{dn_s}{d(\ln k)}\ln\left(\frac{k}{k_*}\right) + \dots\right) \ln\left(\frac{k}{k_*}\right),\\
\ln\left[ P_t(k)\right] =&~\ln(A_t) +\left(n_t + \frac{1}{2}\frac{dn_t}{d(\ln k)} \ln\left(\frac{k}{k_*}\right) + \dots\right)\ln\left(\frac{k}{k_*}\right),
\end{split}
\end{equation}
where $A_s~(A_t)$ denotes the scalar (tensor) amplitude, $n_s~(n_t)$ the scalar (tensor) spectral index, $k_*$ the mode $k$ crossing the Hubble radius, and $\mathcal{R}$ the comoving curvature perturbation, which is approximately equal to the comoving curvature perturbation on uniform density hypersurfaces, $\zeta$, on large scales. Henceforth we will focus on the scalar part of the power spectrum. For considering the pure power-law form of Eq.\ (\ref{powerdef}), $P_\mathcal{R}(k) = A_s \left(k/k_*\right)^{n_s -1}$, the \textit{Planck} collaboration \cite{2013arXiv1303.5082P} recently found the best fit values $A_s = 2.2\times10^{-9}$ and $n_s = 0.9603~(\pm 0.0073)$ for $k_*=0.05$ Mpc$^{-1}$, which constrain all inflationary scenarios. However, also an extension to this simple power-law shape is currently investigated taking into account bumps, sharp features, or wiggles. These types of deviations are well motivated by, e.g., implications of the recent BICEP2 data \cite{2014arXiv1403.3985B,2014arXiv1403.7786H,2014arXiv1404.3690H,2014arXiv1405.2012H}, or special features of the inflaton potential \cite{2014arXiv1404.2985E,2014arXiv1404.6093M}. Such features, in turn, might indicate non-linear physics and thus correspond to non-vanishing non-Gaussianity parameters \cite{2013arXiv1303.5082P}. These features would therefore be an indicator for inflation models beyond single-field slow-roll scenarios.~$\blacktriangleleft$

\medskip
The primordial power spectrum is a valuable quantity since it depends on the physics of the early Universe. Its inference process is highly non-trivial. Therefore we would like to present two Bayesian, non-parametric reconstruction schemes in the framework of information field theory, which, however, are related to each other. To reconstruct the primordial power spectrum we have to know the inferred field $\zeta$ and its variance. This means we are interested in the posterior pdf of $\zeta$. Following Sec.\ \ref{sec:ift} the Hamiltonian is given by
\begin{equation}
\label{recon}
H(d,\zeta) = -\ln\left[\mathcal{G}(d-R\zeta,N)P(\zeta)\right].
\end{equation}
In general, Eq.\ (\ref{recon}) cannot be evaluated further because the shape of $P(\zeta)$ is unknown. Fortunately, a Gaussian is a very good approximation for $P(\zeta)$ as argued in Sec.\ \ref{sec:postder} and motivated by the actual constraints on $f_\text{NL}$ \cite{2013arXiv1303.5084P} that becomes exact if $f_\text{NL}=g_\text{NL}=0$. This enormously simplifies the derivation and under this approximation the posterior is given by
\begin{equation}
P(\zeta|d) = \mathcal{G}(\zeta - m_\text{w},D),
\end{equation}
with $m_\text{w} \equiv  Dj = \left(\Xi^{-1} + M\right)^{-1}R^\dag N^{-1}d$ the Wiener filter solution.

\medskip
One may also be interested in the case\footnote{We declare the case of $|f_\text{NL}|\propto \mathcal{O}(1)$ to be interesting due to the current constraints on $f_\text{NL}$. The discussion that follows, however, is generic and therefore valid for arbitrary values of $f_\text{NL}$ and $g_\text{NL}$ still satisfying the saddle-point approximation, Eq.\ (\ref{postgen}).} where $|f_\text{NL}|\propto \mathcal{O}(1)$ and $g_\text{NL}\neq 0$. Here, the quantity of interest is the power spectrum of the primordial Gaussian perturbation $\zeta_1$, $P_{\zeta_1}(k)$. The approach of reconstructing $\zeta_1$ for fixed non-Gaussianity parameters is already described in Sec.\ \ref{sec:postder} and determined by Eqs.\ (\ref{gradgen}) and (\ref{hessgen}). Therefore the posterior is given by $\mathcal{G}(\zeta_1 - \bar{\zeta_1}, D_{d,f_\text{NL},g_\text{NL}})$, which implicitly depends on parameters of non-Gaussianity or alternatively on parameters of inflation.

\medskip
Given a reconstructed map of $\zeta$ or $\zeta_1$ and its uncertainty $D$ or $D_{d,f_\text{NL},g_\text{NL}}$, the challenge is now to appropriately infer the power spectrum $P_\zeta(k)$ or $P_{\zeta_1}(k)$ under consideration of the uncertainty. For this purpose we suggest the two following approaches (Eqs.\ (\ref{crit}) and (\ref{smooth})), which have already successfull applications in cosmology and astrophysics, e.g., Refs.\ \cite{2010PhRvE..82e1112E,2012A&A...542A..93O,2013arXiv1311.5282J,2013arXiv1311.1888S}. We will show that these methods are able to reconstruct the spectrum of the primordial curvature perturbations even in case of significant non-Gaussianity and partial sky coverage.

\subsection{Filter formulae}
\textbf{Critical filter.} The first filter captures the concepts of the well known Karhunen-Lo\`{e}ve \cite{Kar,Loeve1978} and Feldman-Kaiser-Peacock \cite{1994ApJ...426...23F} estimators and has been derived in Ref.\ \cite{2011PhRvD..83j5014E}. The aim here is to reconstruct the power spectrum for Gaussian signals, which determines the statistics completely under the cosmological assumption of translationally and rotationally invariance. This implies the existence of an orthonormal basis $O$ in which $\Xi$ become diagonal, e.g., the Fourier space with elements $\vec{k}=(k_1,\dots,k_3)\in\mathds{R}^3$ of length $k\equiv|\vec{k}|$ (Fourier mode) for signals defined in Euclidean space, or the spherical harmonics space for signals defined on the sphere. Following Ref.\ \cite{2011PhRvD..83j5014E}, the signal covariance, here $\Xi_{kk'}$, and its inverse are linearly parametrized by non-overlapping basis functions $f_i(k)$, commonly denoted as spectral bands, and coefficients $\tilde{p}$,
\begin{equation}
P_\zeta(k) = \sum_i \tilde{p}_i f_i(k),
\end{equation}
where $\left(\Xi_i\right)_{xy} = O^*_{xk} f_i(k) O_{ky}$, and therefore
\begin{equation}
\Xi_{\tilde{p}} = \sum_i \tilde{p}_i ~\Xi_i ~~\text{and}~~\Xi^{-1}_{\tilde{p}} = \sum_i\tilde{p}^{-1}_i ~\Xi^{-1}_i.
\end{equation}
$\Xi^{-1}_i$ denotes the pseudo-inverse of the band-variances, given by $\left(\Xi^{-1}_i\right)_{xy} = O^*_{xk} g_i(k) O_{ky}$ with $g_i(k)= 1/f_i(k)$ if $f_i(k)>0$ and $g_i(k)=0$ if $f_i(k) =0$ \cite{2011PhRvD..83j5014E}. In all cases addressed in this paper the non-overlapping basis functions $f_i(k)$ are projections from the vectors $\vec{k}$ onto the Fourier modes $k=|\vec{k}|=\sqrt{k_1^2+k_2^2 + k_3^2}$. This way, the primordial power spectrum can be parametrized as in Eq.\ (\ref{pex}). 

The priors of $\tilde{p}$ are assumed to be mutually independent, $P(\tilde{p}) = \prod_i P(\tilde{p}_i)$, and obey an inverse Gamma distribution,
\begin{equation}
P(\tilde{p}_i) = \mathcal{I}(\tilde{p}_i; \alpha_i, q_i) \equiv  \frac{1}{q_i\Gamma(\alpha_i -1)}\left(\frac{\tilde{p}_i}{\alpha_i}\right)^{-\alpha_i}\exp{\left(-\frac{q_i}{\tilde{p}_i} \right)},
\end{equation}
with $\Gamma$ the Gamma function. Constructed in this way one obtains an informative prior by $\alpha_i \gg 1$ and a non-informative prior, e.g.\ Jeffreys prior, by $\alpha_i = 1, q_i =0$.

To derive the critical filter formula we calculate the minimum of the $\zeta$ marginalized Hamiltonian, 
\begin{equation}
H(d,\tilde{p}) = \frac{1}{2}\text{tr}\left(\ln \Xi \right) -\frac{1}{2}\text{tr}\left(\ln D \right) -\frac{1}{2}j^\dag D j + \sum_i (\alpha_i -1)\tau_i + q_i e^{-\tau_i}, 
\end{equation}
with respect to $\tau \equiv  \ln(\tilde{p})$, thereby maximizing the posterior probability for the logarithmic power spectrum yielding the coupled system of equations \cite{2011PhRvD..83j5014E}
\begin{equation}
\label{crit}
\begin{split}
m_{\tilde{p}_{\text{min}}} =&~ D_{\tilde{p}_{\text{min}}}j,\\
\tilde{p}_{i,\text{min}} =&~ \frac{q_i + \frac{1}{2}\text{tr}\left(m_{\tilde{p}_{\text{min}}}m^\dag_{\tilde{p}_{\text{min}}} + D_{\tilde{p}_{\text{min}}} \right)\Xi^{-1}_i}{\alpha_i -1 +\frac{1}{2}\text{tr}\left(\Xi^{-1}_i \Xi_i \right)}.
\end{split}
\end{equation}
For the parameter choice according to Jefferys prior, $\alpha_i = 1, q_i =0$, Eq.\ (\ref{crit}) is called critical filter. To solve this coupled system iteratively, we need a boundary condition, e.g., for the power spectrum (remember that $D_{\tilde{p}_{\text{min}}}$ depends on the spectral coefficients $\tilde{p}_{i,\text{min}}$). A well motivated initial guess might be the primordial power spectrum from Planck \cite{2013arXiv1303.5082P}, which is a pure power law, Eq.\ (\ref{powerdef}).

\medskip
\textbf{Critical filter with smoothness prior.} For some physical reasons \cite{2013arXiv1303.5082P}, e.g., that physics do not change suddenly during inflation, one may want to enforce the reconstructed power spectrum to be smooth. This can be incorporated by an extension of the prior \cite{2013PhRvE..87c2136O}, given by
\begin{equation}
P(\tau) =P_{\text{sm}}(\tau) \prod_i P(\tau_i),
\end{equation}
with the smoothness prior
\begin{equation}
P_{\text{sm}}(\tau)\propto \exp{\left(-\frac{1}{2\sigma_\tau^2}\int d(\ln k) \left(\frac{\partial^2\ln \tilde{p}(\tau_k)}{\partial (\ln k)^2} \right)^2 \right)}\equiv \exp{\left(-\frac{1}{2}\tau^\dag T \tau \right)},
\end{equation}
which punishes any deviation from a power-law power spectrum with a strength $\sigma_\tau$. This means in the limit of $\sigma_\tau\rightarrow \infty$ we recover Eq.\ (\ref{crit}) whereas for a finite decreasing, especially small value of $\sigma_\tau$ the smoothness increases. Here we introduced the linear operator $T$ whose explicit form can be found in Ref.\ \cite{2013PhRvE..87c2136O}. $T$ includes the integral as well as the scaling constant $\sigma_\tau$. An analogous derivation to the critical filter case then yields
\begin{equation}
\label{smooth}
\tilde{p}_{i,\text{min}} = \frac{q_i + \frac{1}{2}\text{tr}\left(m_{\tilde{p}_{\text{min}}}m^\dag_{\tilde{p}_{\text{min}}} + D_{\tilde{p}_{\text{min}}} \right)\Xi^{-1}_i}{\alpha_i -1 +\frac{1}{2}\text{tr}\left(\Xi^{-1}_i \Xi_i \right) + \left(T\tau \right)_i}.
\end{equation}
In comparison to Eq.\ (\ref{crit}) the result exhibits the additional term $(T\tau)_i$ in the denominator that enforces smoothness. By appropriately choosing the scale parameter $\sigma_\tau$ one is able to permit the reconstruction of features on specific scales. However, for a detailed discussion of the critical filter with smoothness prior including the choice of $\sigma_\tau$ see Ref.\ \cite{2013PhRvE..87c2136O}.

\subsection{Numerical toy example}\label{sec:toy}
\subsubsection{Inferring a power spectrum of approximately Gaussian curvature perturbations}
\begin{figure}[ht]
\begin{center}
\includegraphics[width=.5\textwidth]{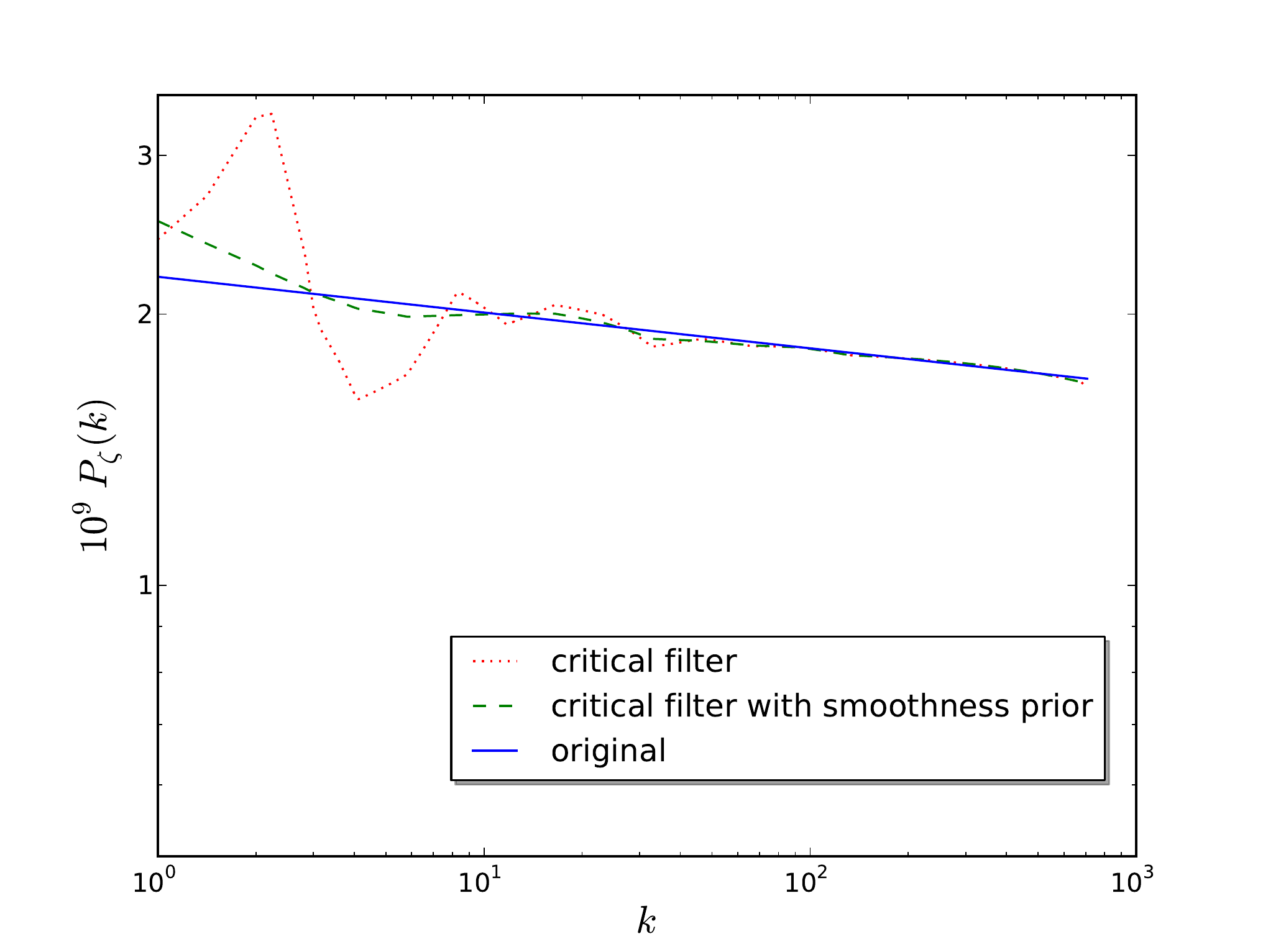}%
\includegraphics[width=.5\textwidth]{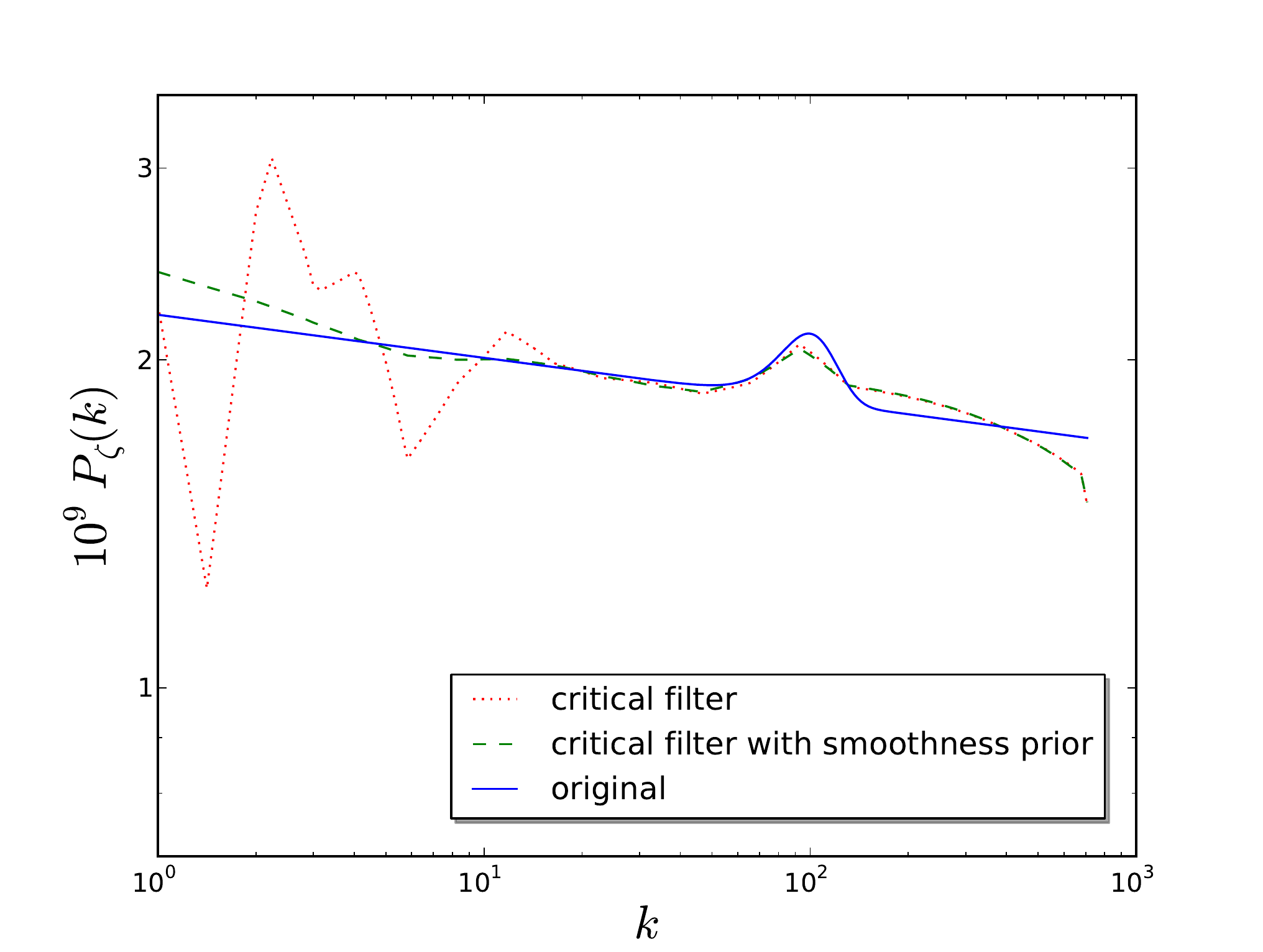}\\
(a) \hspace{7cm} (b)\\

\includegraphics[width=.5\textwidth]{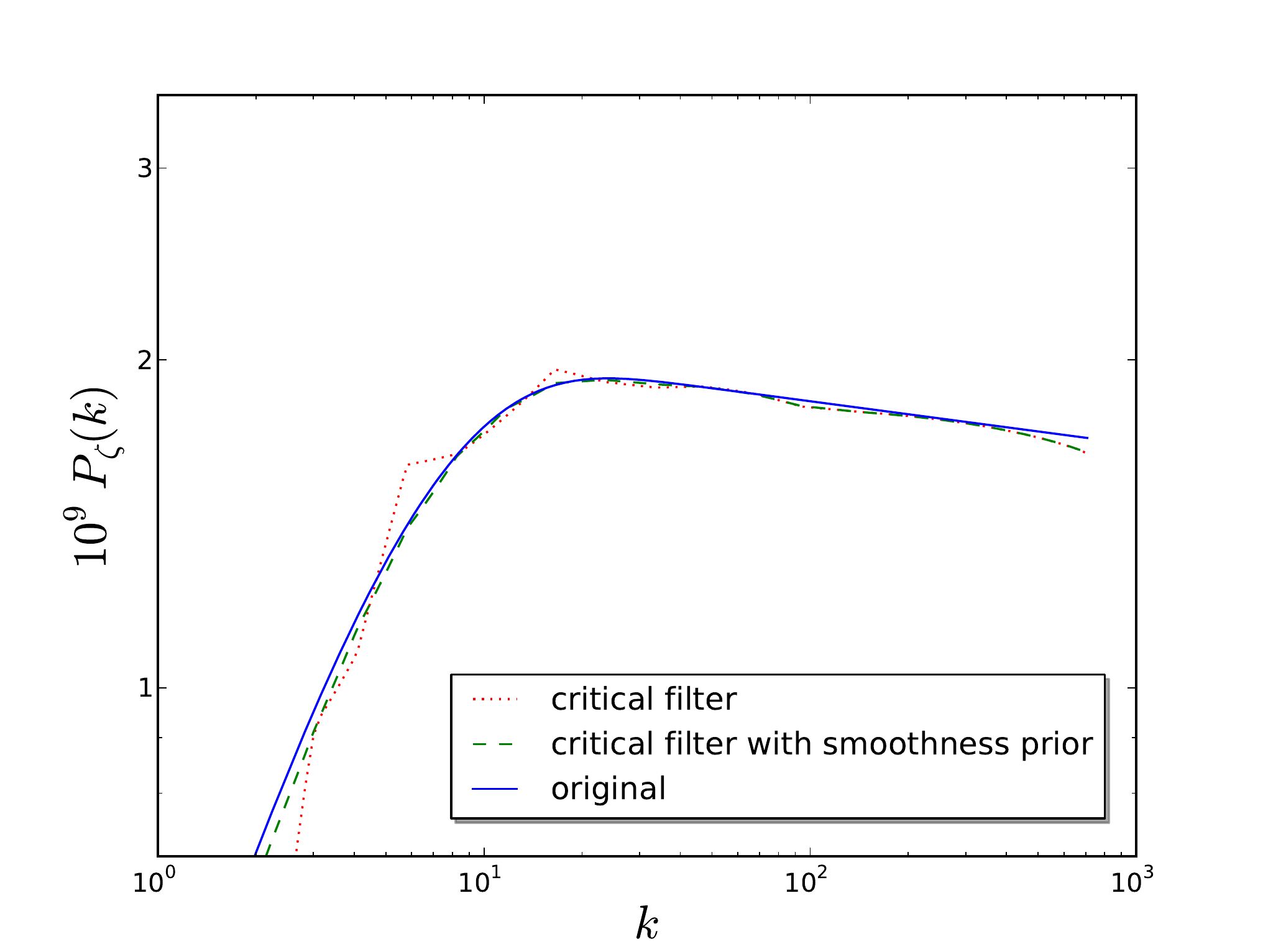}\\
(c)
\end{center}
\caption[width=\columnwidth]{(Color online) Primordial power spectrum reconstruction of approximately Gaussian curvature perturbations without (a) and with [(b), (c)] features by a critical filter solely (red dotted line) and including a smoothness prior (green dashed line) according to Eqs.\ (\ref{crit}) and (\ref{smooth}) and compared to the original power spectrum (blue solid line).}
\label{pps}
\end{figure}

To demonstrate the performance of the filter formulae, Eqs.\ (\ref{crit}) and (\ref{smooth}), according to the inference of a power spectrum of approximately Gaussian curvature perturbations $\zeta$ we use the two dimensional test case of Sec.\ \ref{sec:impl}, but with $N_\text{pix} = 10^6$, and $f_\text{NL},g_\text{NL}\approx 0$ to satisfy the condition of negligible non-Gaussianity. Additionally, we adopt the parameter choice according to Jefferys prior, i.e.\ $\alpha_i = 1, q_i =0$, and a scaling constant of $\sigma^2_\tau = 0.1$. Figure \ref{pps}~(a) shows the result and confirms the properness of the reconstruction, which exhibits typical deviations from the true underlying spectrum for low $k$ modes due to the effect of cosmic variance. Reconstruction errors are not included in the figure but could be incorporated by evaluating the inverse Hessian of the Hamiltonian $H(d,\tilde{p})$ for real scenarios, as done in Ref.\ \cite{2013PhRvE..87c2136O}. 

\begin{figure}[ht]
\begin{center}
\includegraphics[width=.4\textwidth, height = 5.6cm]{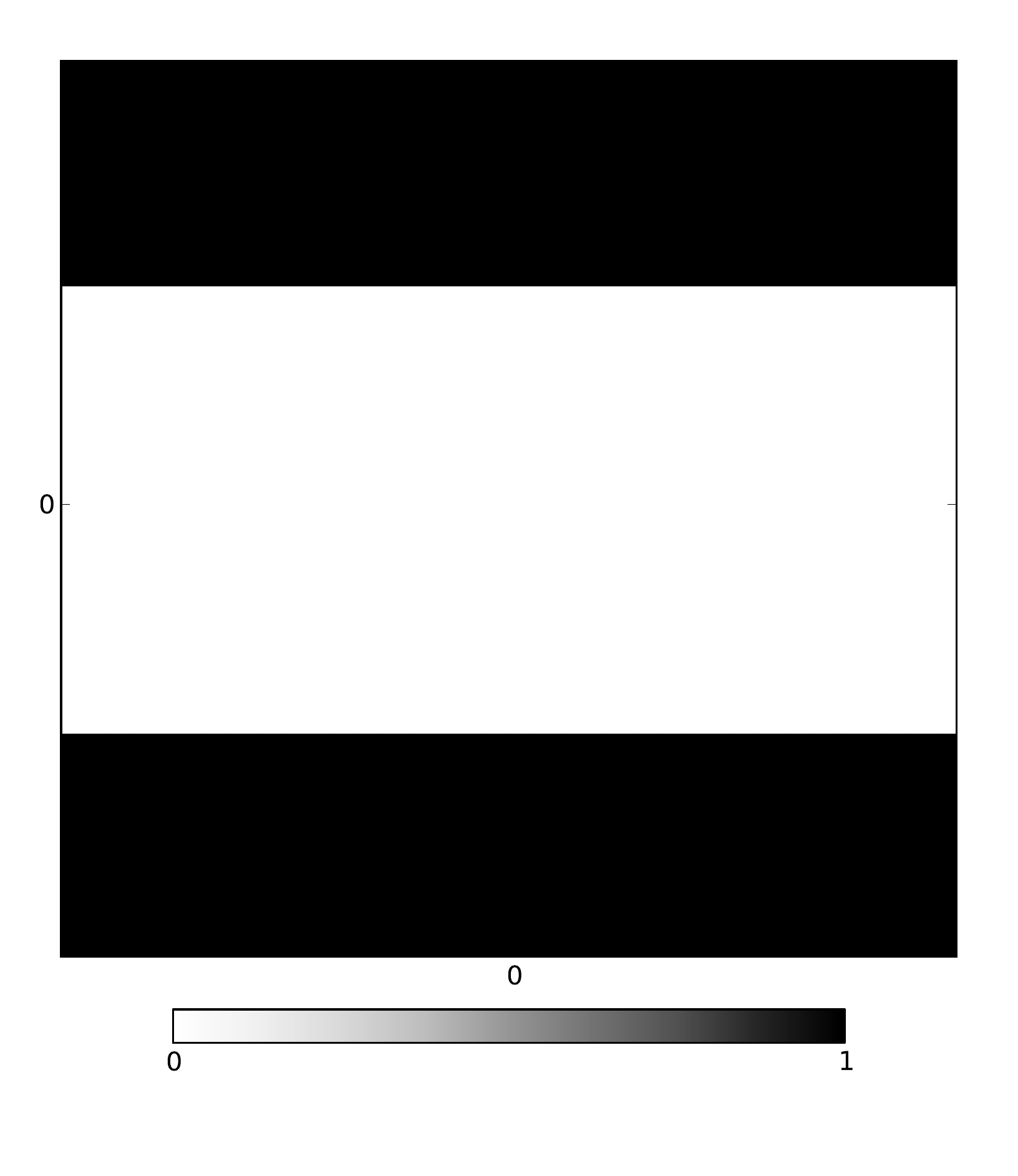}%
\includegraphics[width=.5\textwidth]{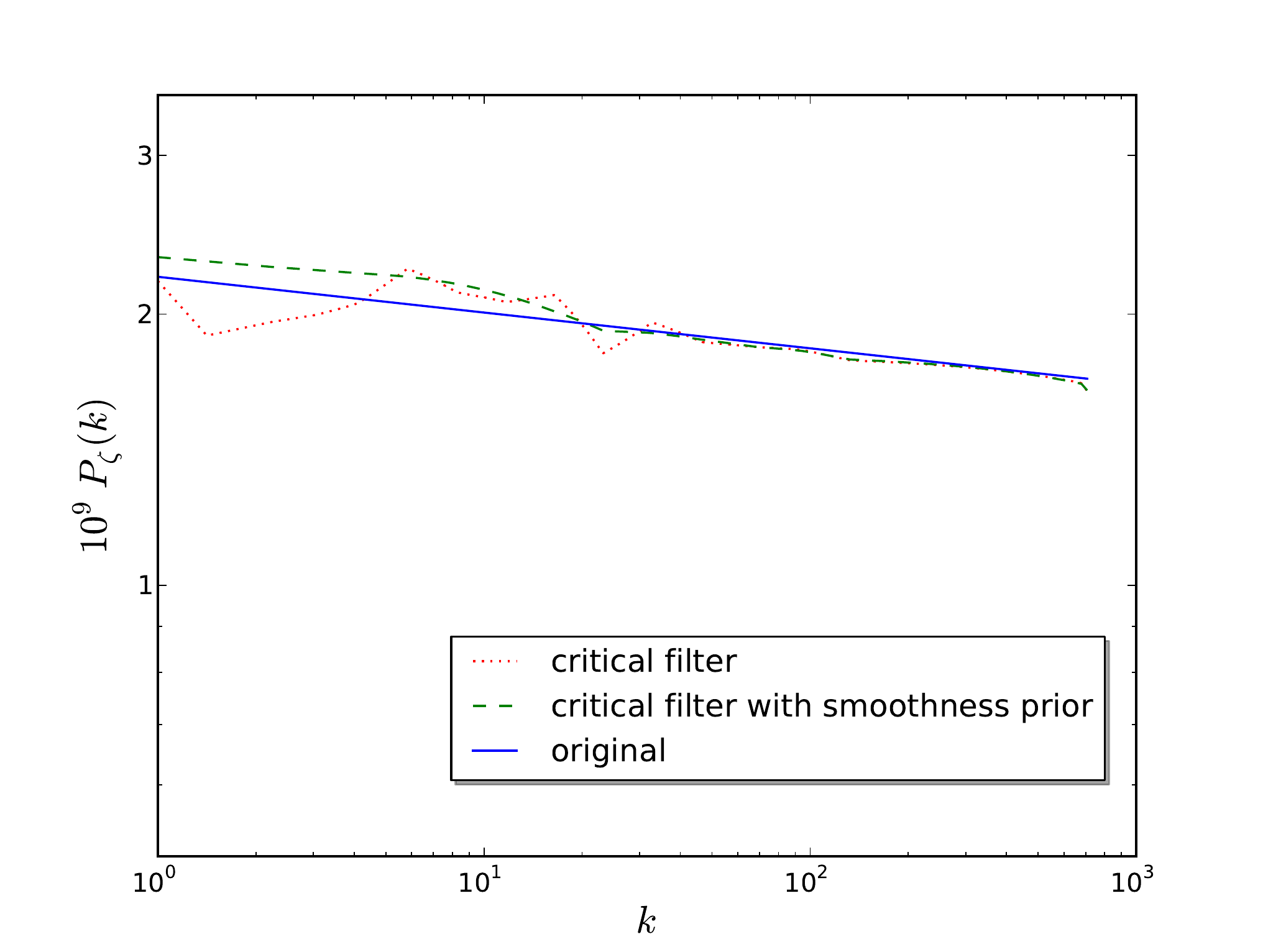}\\
(a) \hspace{6cm} (b)
\end{center}
\caption[width=\columnwidth]{(Color online) Primordial power spectrum reconstruction of approximately Gaussian curvature perturbations [see (b)] at a sky coverage of $50\%$ [see (a)] by a critical filter solely (red dotted line) and including a smoothness prior (green dashed line) according to Eqs.\ (\ref{crit}) and (\ref{smooth}) and compared to the original power spectrum (blue solid line).}
\label{pps_sky}
\end{figure}

The ability of a non-parametric reconstruction of features on the power spectrum like bumps and cutoffs as well as dealing with a partial sky coverage is also illustrated by Figs.\ \ref{pps}~[(b),(c)] and \ref{pps_sky}. For the latter case we consider a mask in addition to the response, $R_\text{mask}$, so that the sky is observed by $50\%$ only. In all cases the critical filter with and without smoothness prior works well, i.e., it is able to reconstruct the primordial power spectrum including possible features.

\begin{figure}[ht]
\begin{center}
\includegraphics[width=.5\textwidth]{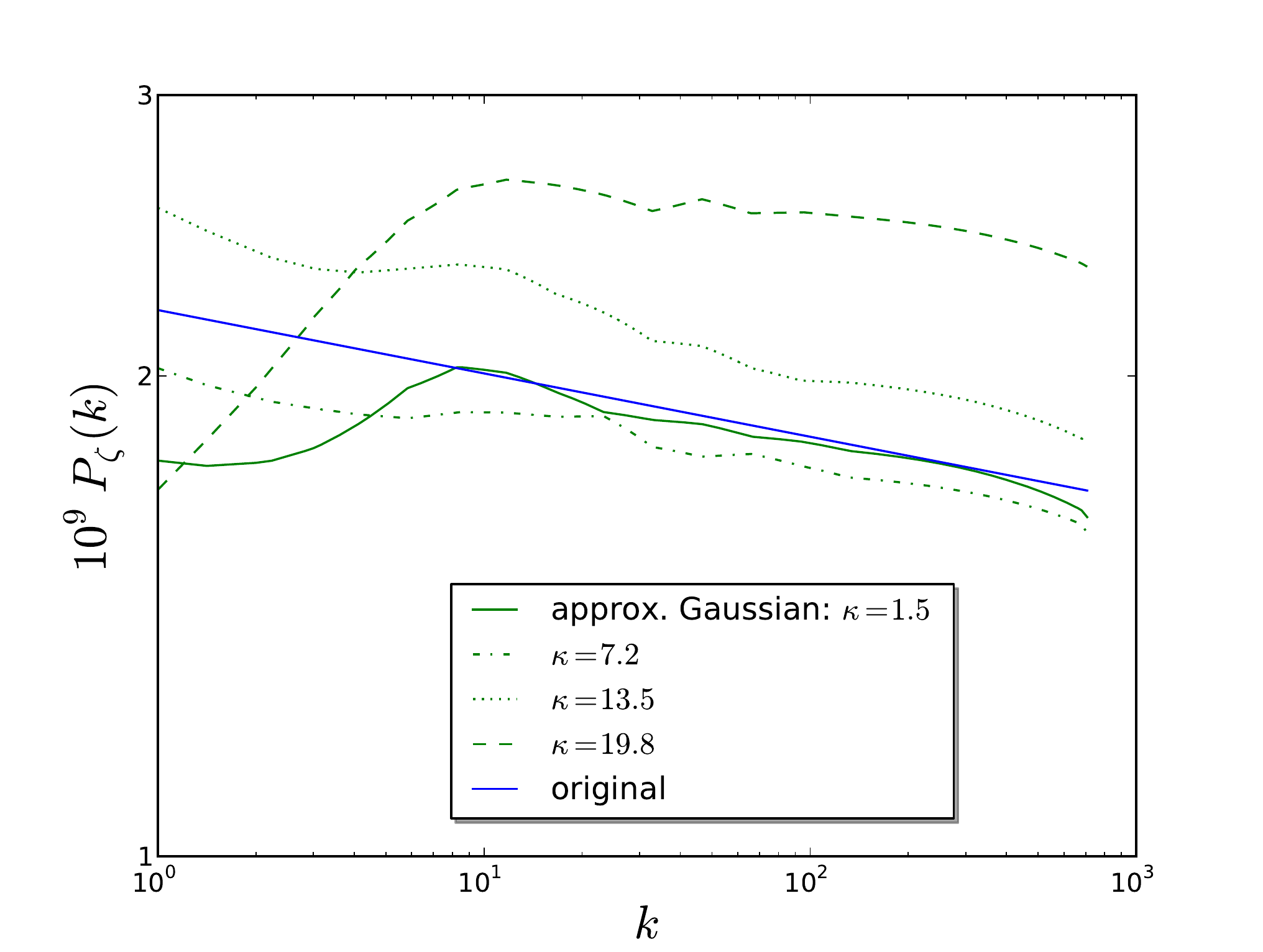}%
\includegraphics[width=.5\textwidth]{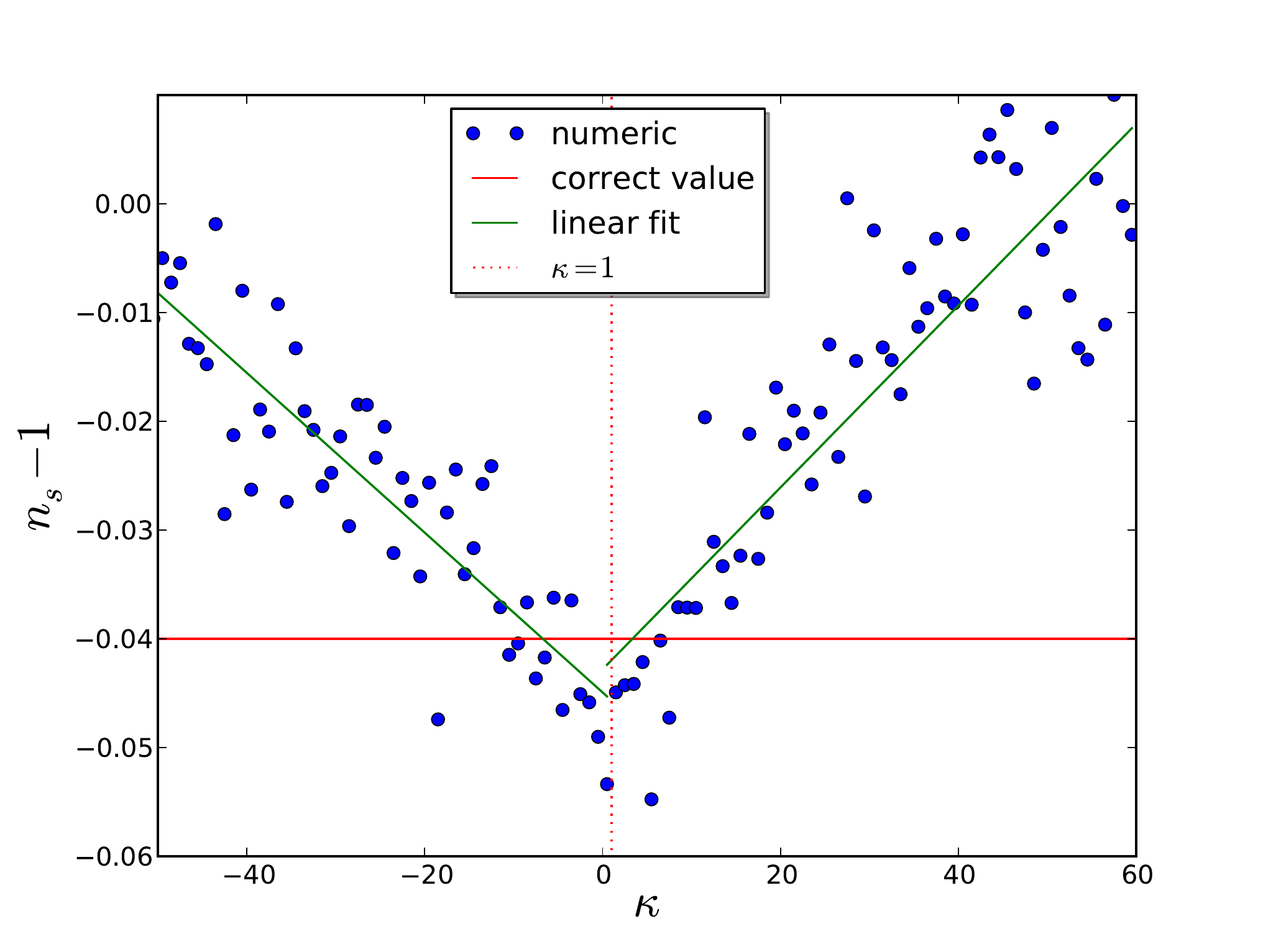}\\
(a) \hspace{7cm} (b)\\
\includegraphics[width=.5\textwidth]{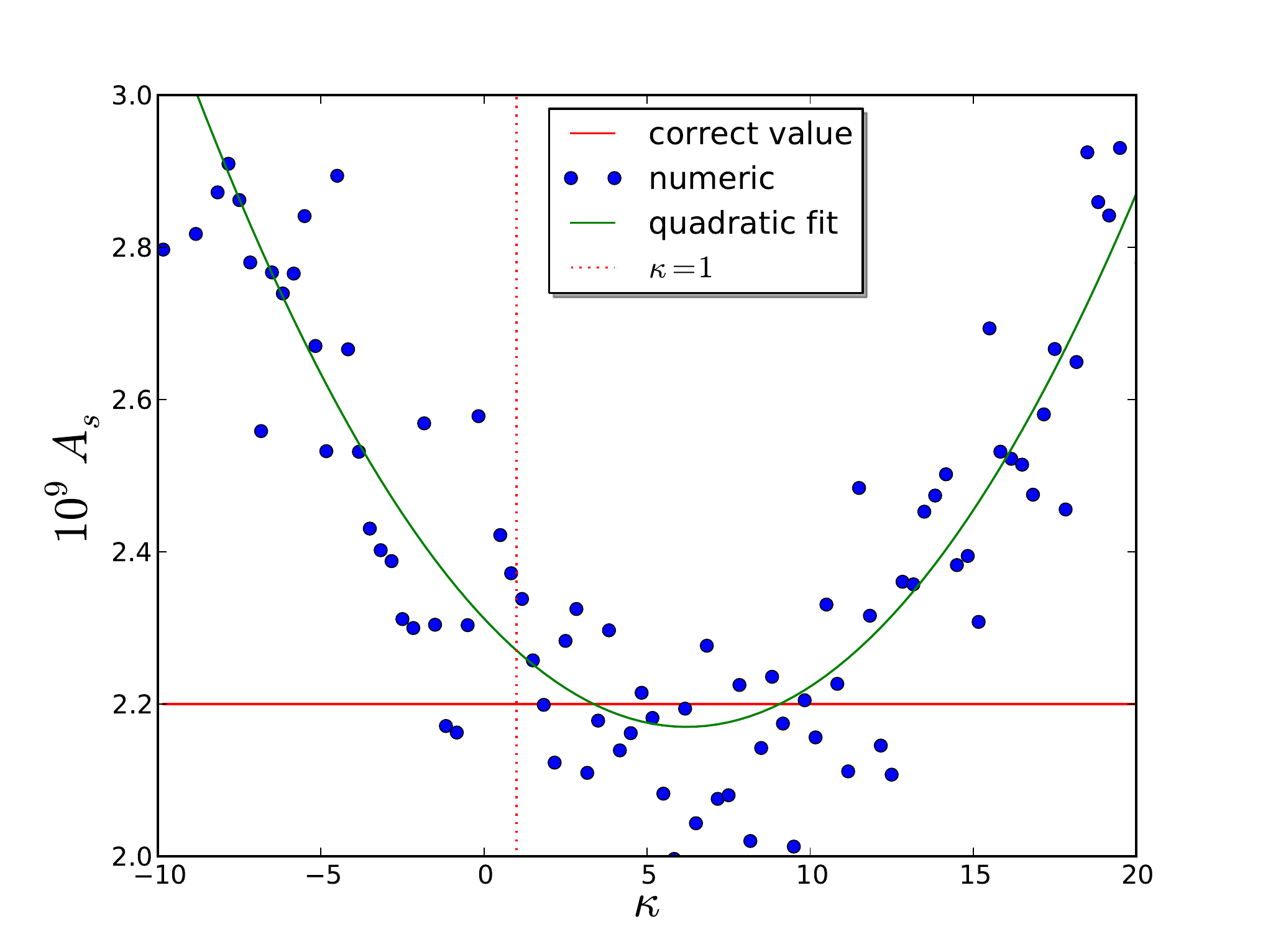}%
\includegraphics[width=.5\textwidth]{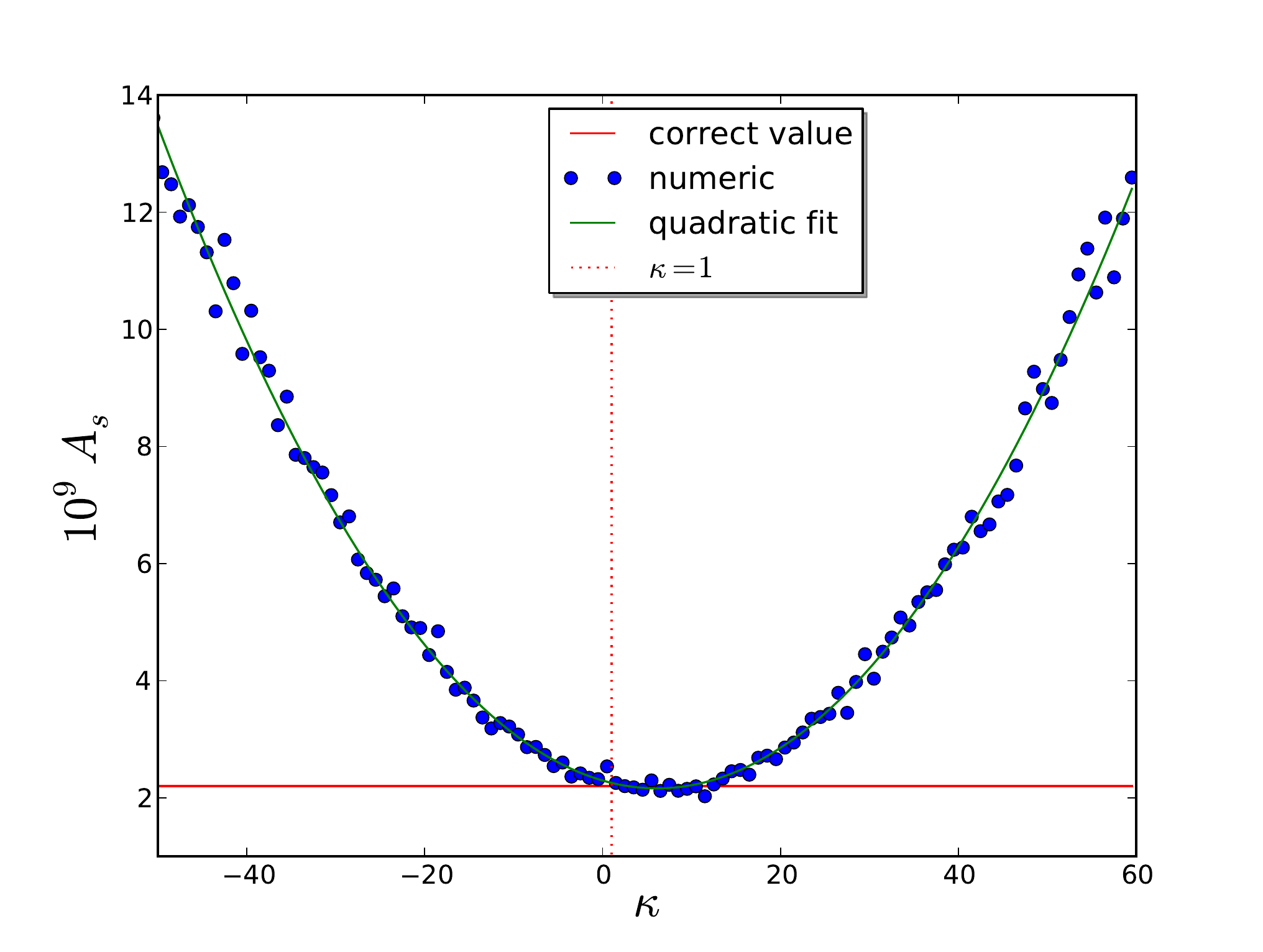}\\
(c) \hspace{7cm} (d)
\end{center}
\caption[width=\columnwidth]{(Color online) Primordial power spectrum reconstruction of non-Gaussian curvature perturbations, which are falsely assumed to be Gaussian, in the curvaton scenario for various choices of $\kappa$ by a critical filter solely and including a smoothness prior according to Eqs.\ (\ref{crit}) and (\ref{smooth}). Panel (a) shows the comparison to the original power spectrum. The dependence of $A_s$ [(c), (d)] and $n_s -1$ [(b)] on $\kappa$ has been determined by conducting a linear fit to the reconstructed power spectrum for $k\gtrsim 100$.}
\label{scaling}
\end{figure}

We want to emphasize, however, that for a proper reconstruction the condition of Gaussianity has to be fulfilled because the approach introduced is very sensitive to deviations from this restriction. To illustrate this we consider again the curvaton scenario where $f_\text{NL}$ and $g_\text{NL}$ are parametrized as a function of $\kappa$, cf.\ Eq.\ (\ref{fg_curv}). This means that for large values of $\kappa$ the curvature perturbation $\zeta$ is now falsely assumed to be Gaussian with deviations according to Eq.\ (\ref{fg_curv}). Fig.\ \ref{scaling} shows the performance of the critical filter with smoothness prior for this case. Panel (a) shows how the reconstructed power spectrum deviates from the true underlying one as a function of the level of non-Gaussianity, parametrized by $\kappa$. For a small level of non-Gaussianity there is no observable effect on the spectral index $n_s$ whereas the scalar amplitude $A_s$ depends strongly on $\kappa$ as shown by panel (c) and (d). The quadratic fits appearing within this panels obey the formula $10^9A_s(\kappa) = a_q\kappa^2 +b_q\kappa + c_q$ with $a_q=0.0036,~b_q=- 0.044$ and $c_q=2.3$ for $\kappa \geq 1$, which can be reformulated analytically into a $f_\text{NL}$ dependency, if we neglect contributions of the trispectrum (see App.\ \ref{app2}).

For a higher level of non-Gaussianity also $n_s$ becomes affected [Fig.\ \ref{scaling} (b)]. The two linear fits in panel (b) are generated independently, because the region of $\kappa<1$ is unphysical. The physically relevant fit for $\kappa\geq 1$ obeys the formula $n_s(\kappa) -1 = a_\mathrm{l}\kappa +b_\mathrm{l}$ with $a_\mathrm{l}=8\times10^{-4}$ and $b_\mathrm{l}=-0.04$. The latter formula can also be reformulated into a dependency on $f_\text{NL}$. Additionally one can derive a quadratic relation between the spectral index and the scalar amplitude, cf.\ App.\ \ref{app2}. 

Note that these particular dependencies on the level of non-Gaussianity (or alternatively on inflationary parameters) is not a generic statement, but valid for the critical filter with and without smoothness prior, which were not informed here about the presence of non-linearities, and that the magnitude of the deviations depends additionally on the number of pixels used.
\subsubsection{Inferring a power spectrum of non-Gaussian curvature perturbations}
\begin{figure}[ht]
\begin{center}
\includegraphics[width=.5\textwidth]{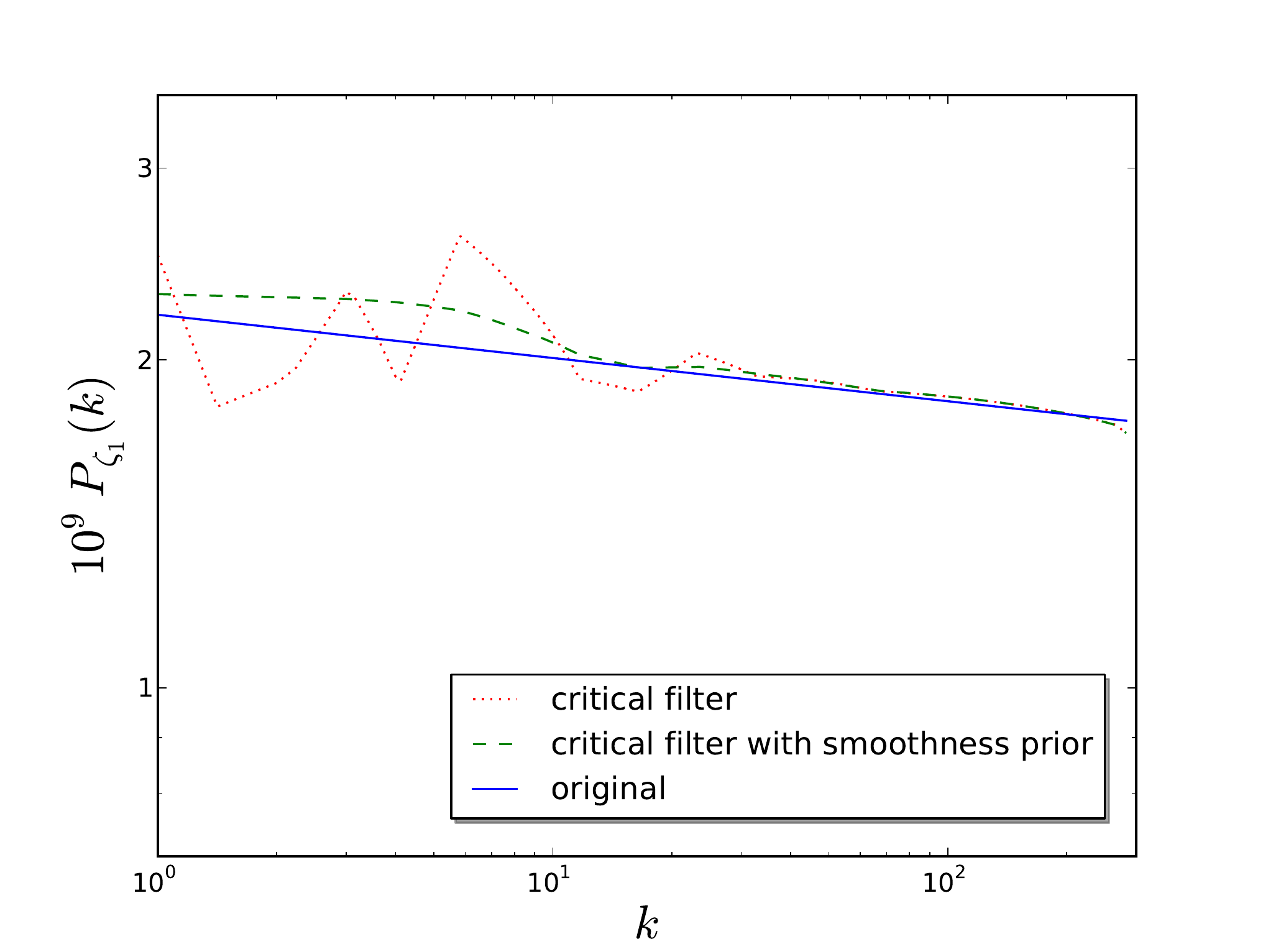}%
\includegraphics[width=.5\textwidth]{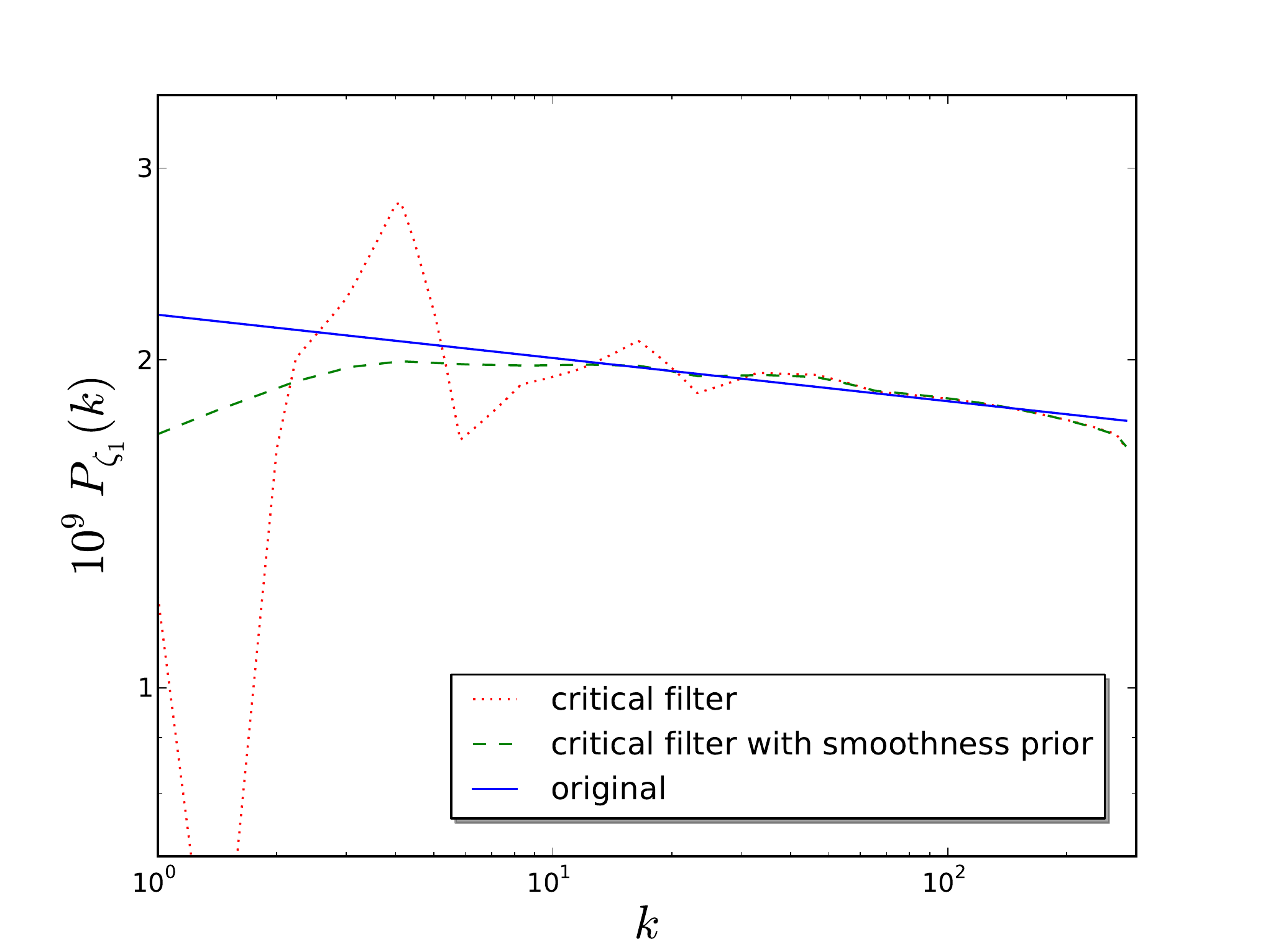}\\
(a) \hspace{7cm} (b)
\end{center}
\caption[width=\columnwidth]{(Color online) Primordial power spectrum reconstruction of non-Gaussian curvature perturbations with non-vanishing [(a) $\kappa_\text{gen} = 7.2$, (b) $\kappa_\text{gen} = 13.5$] non-Gaussianity by a critical filter solely (red dotted line) and including a smoothness prior (green dashed line) according to Eqs.\ (\ref{crit}) and (\ref{smooth}) and compared to the original power spectrum (blue solid line).}
\label{pps_NG}
\end{figure}

Now we show that the critical filter with and without smoothness prior in combination with results of Sec.\ \ref{sec:ift} is able to reconstruct the spectrum of the Gaussian, primordial curvature perturbations $\zeta_1$ even in case of significant non-Gaussianity.
For this case (of inferring a power spectrum of non-Gaussian curvature perturbations $\zeta$) we leave all numerical specifications in place, but use $N_\text{pix}=1.6\times 10^5,~P_{\zeta_1}(k)\delta_{kk'}$ according to Eq.\ (\ref{powerdef}), and use $\kappa=7.2~(\widehat{=}~f_\text{NL},g_\text{NL}=7.2,-29),~13.5~(\widehat{=}~f_\text{NL},g_\text{NL}=15,-56)$ to seed a non-vanishing level of non-Gaussianity. The reconstructed map of $\zeta_1$, from which we infer the power spectrum, is calculated according to Eq.\ (\ref{gradgen}) and its uncertainty according to Eq.\ (\ref{hessgen}). Note that these quantities now depend on $f_\text{NL}(p)$ and $g_\text{NL}(p)$ and thus on a specific inflation model. Figure \ref{pps_NG} shows the result both for critical filter and its extension including a smoothness prior. Reconstruction errors are also not included. Compared with the improper reconstructed power spectra of Fig.\ \ref{scaling} (a) (compare in particular the case of $\kappa=13.5$), the advanced method used here yields adequate results. This comparison also suggests that one could infer the level of non-Gaussianity by measuring both, non-Gaussian and Gaussian power spectrum. $\blacktriangleright$ These spectra might be inferred from, e.g., $T-$ and $B-$modes due to the fact that $B-$modes might be less non-Gaussian than $T-$ modes \cite{2010PhRvD..82j3505S}. Afterwards the level of non-Gaussianity could be determined by the difference between the respective spectral amplitudes.~$\blacktriangleleft$ 

Once the power spectrum of the Gaussian curvature perturbation $\zeta_1$ is determined the power spectrum of the non-Gaussian curvature perturbation, $\left\langle \zeta \zeta^\dag \right \rangle_{P(\zeta)}$, can also be calculated from $P_{\zeta_1}(k)$. The distribution $P(\zeta)$ required for this calculation can be calculated approximately and is pointed out, e.g., in Ref.~\cite{2005PhRvD..72d3003B}.

\section{Conclusion}\label{sec:conclusion}
We have presented a novel and generic method to infer inflation models from observations by the non-Gaussianity parameters $f_\text{NL}$ and $g_\text{NL}$ and how to reformulate this method to infer specific parameters of inflation models, $p$, directly (see especially Secs.\ \ref{sec:ift} and \ref{sec:special_models}). This approach, i.e.\ the analytical derivation of a posterior for $f_\text{NL}$ and $g_\text{NL}$ as well as for $p$ can be used to further distinguish between the already restricted amount of inflation models. It is formulated in a generic manner in the framework of information field theory, so that it is applicable to CMB data as well as to LSS data (see especially the three dimensional example of Sec.\ \ref{sec:scmexp}) by tuning the response appropriately. The analyticity of the method, achieved by a saddle-point approximation, allows to dispense with numerically expensive sampling techniques like the commonly used Monte Carlo method. The analytic approximation we introduced has been validated successfully by the DIP test \cite{paper2}.
 
The second quantity of interest here is the primordial power spectrum due to its farreaching implications for inflationary cosmology. We have presented two computationally inexpensive, approximative Bayesian methods to infer the primordial power spectrum from CMB data, the so called critical filter, Eq.\ (\ref{crit}), and an extension thereof with smoothness prior, Eq.\ (\ref{smooth}). Both methods allow a non-parametric reconstruction of the power spectrum including the reconstruction of possible features on specific scales. Additionally, both methods are able to perform this inference process even in the case of partial sky coverage and non-Gaussianity. We have argued that this property would allow to infer the level of non-Gaussianity of a field if one could measure both, the power spectrum of the non-Gaussian and Gaussian curvature perturbations. These spectra might be inferred from, e.g., $T-$ and $B-$modes due to the fact that $B-$ modes might be less non-Gaussian than $T-$ modes \cite{2010PhRvD..82j3505S}. A fully quantitative analysis thereof, however, is left for future work.

\begin{acknowledgments}
We gratefully acknowledge Eiichiro Komatsu and Vanessa B\"ohm for useful discussions and comments on the manuscript. We also want to thank Marco Selig and Maksim Greiner for numerical support and discussions hereof. KEK would like to thank the Max-Planck-Institute for Astrophysics for hospitality where this work was initiated and acknowledges financial support by Spanish Science Ministry grants FIS2012- 30926 and CSD2007-00042. The work of SH was supported by the DFG cluster of excellence \textit{Origin and Structure of
the Universe} and by TRR 33 \textit{The Dark Universe}. Calculations were realized using the \textsc{NIFTy} \cite{2013arXiv1301.4499S} package\footnote{\url{http://www.mpa-garching.mpg.de/ift/nifty/}}.
\end{acknowledgments}
\appendix
\section{Shape of posterior and estimator of inflationary parameters $p$}\label{app}
\begin{figure}[ht]
\begin{center}
\includegraphics[width=.8\columnwidth]{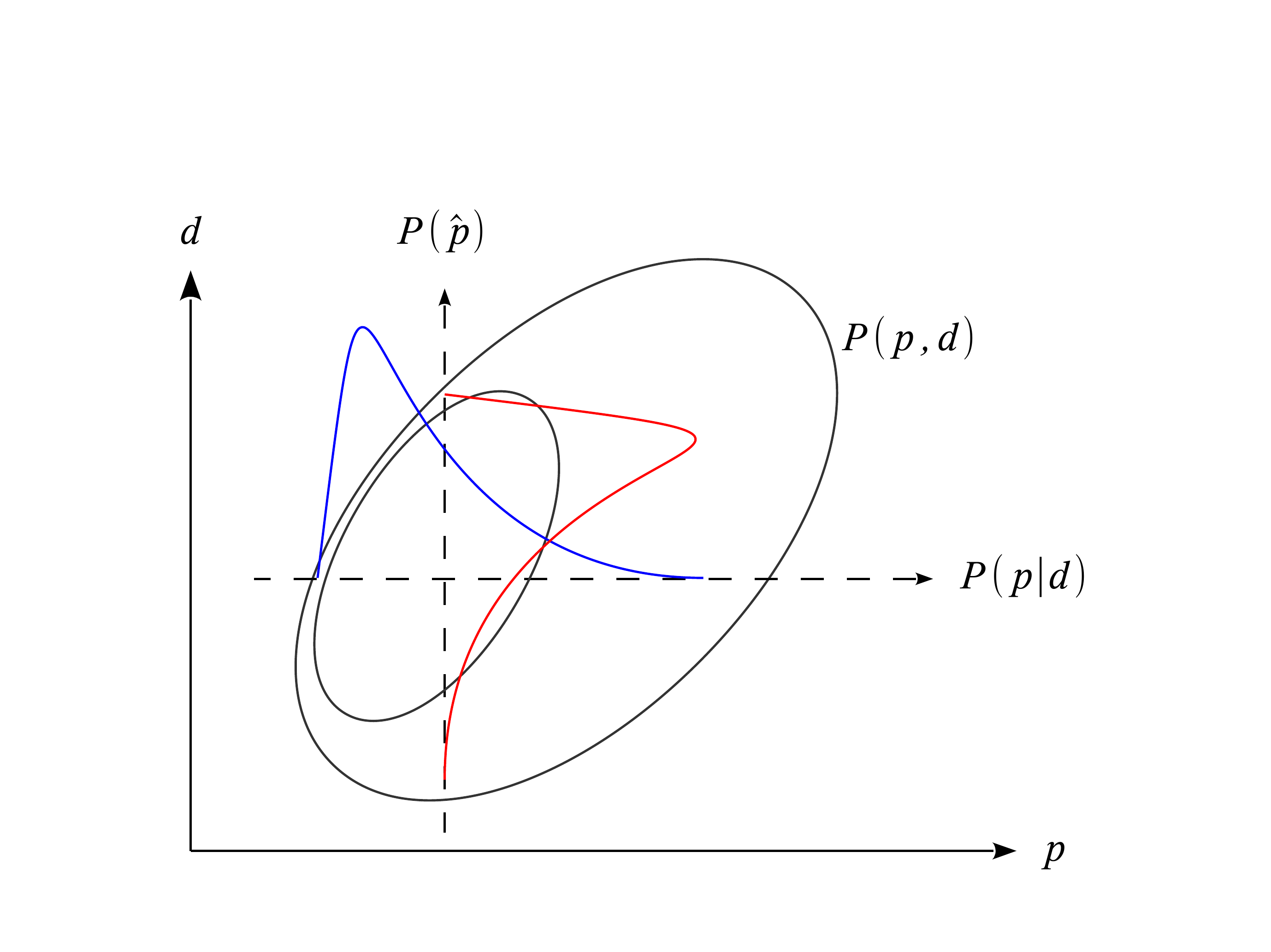}
\end{center}
\caption[width=\columnwidth]{(Color online) Sketch of posterior (blue) and estimator (red) pdf with different skewness behaviors drawn from an asymmetric joint probability (black).}
\label{bias}
\end{figure}

In general, the posterior distribution for $p$ does not have to be Gaussian. If so, one should be very careful if one compares the posterior pdf for $p$ with an estimator pdf, $P(\hat{p})$, because they can exhibit different types of deviations from Gaussianity. This means in particular that in some cases an unbiased constructed estimator might exhibit a skewness behavior different from the posteriors one. For instance, compare the posterior pdf in Ref.\ \cite{paper1} with the estimator pdf in Ref.\ \cite{2011PhRvD..84f3013S}, where the pdf is negatively skewed in one case and positively in another. The reason for this apparent contradiction is illustrated in Fig.\ \ref{bias}, where the joint probability of $p$ and $d$ is shown. To determine the posterior pdf we consider a varying $p$ given $d$, which is a one-dimensional hypersurface, parallel to the horizontal axis. If one wants to obtain the pdf for the estimator, one has to vary $d$ given $p$ corresponding to a one-dimensional hypersurface parallel to the vertical axis. If the probability distribution is symmetric, e.g.\ Gaussian, the shapes of estimator pdf and posterior pdf coincide. However, if the distribution is asymmetric the shapes do not have to coincide as sketched in Fig.\ \ref{bias} for a one-dimensional parameter $p$. In turn, this means that the shapes of posterior and the distribution of an estimator of a quantity do not have to exhibit the same skewness. Hence there is no real contradiction. A discussion about other advantages and disadvantages of the usage of posterior distributions can be found, e.g., in Ref.\ \cite{2013JCAP...06..023V}.

\section{Dependency of the scalar amplitude and spectral index on $f_\text{NL}$ under usage of the critical filter}\label{app2}
In Sec.\ \ref{sec:toy} a relation between the scalar amplitude $A_s$ (spectral index $n_s$) of the primordial power spectrum and the non-Gaussianity parameter $f_\text{NL}$ was mentioned, if one uses the critical filter with or without smoothness prior to reconstruct the power spectrum from a field that is (in some cases) falsely assumed to be Gaussian. In turn, that means by applying these filter formulae, Eqs.\ (\ref{crit}) and (\ref{smooth}), one could infer the level of non-Gaussianity of a field.

Within Sec.\ \ref{sec:toy}  we did this calculation for the curvaton scenario. However, we can transform the dependency of $A_s,~n_s$ on $\kappa$ into a dependency on $f_\text{NL}$ by neglecting contributions of the trispectrum. Solving Eq.\ (\ref{fg_curv}) for $\kappa$ and substituting it within the quadratic and linear fitting formula, pointed out in Sec.\ \ref{sec:toy}, yields 
 

\begin{equation}
\begin{split}
10^9 A_s(f_\text{NL}) =&~ \frac{a_q}{125}\left(10+6f_\text{NL} +\sqrt{2}\sqrt{18f^2_\text{NL} +60f_\text{NL} +125} \right)^2\\
	  &~ + \frac{b_q}{15}\left(10+6f_\text{NL} +\sqrt{2}\sqrt{18f^2_\text{NL} +60f_\text{NL} +125} \right) +c_q,
\end{split}
\end{equation}
and
\begin{equation}
n_s(f_\text{NL})-1 = \frac{a_\mathrm{l}}{15}\left(10+6f_\text{NL} +\sqrt{2}\sqrt{18f^2_\text{NL} +60f_\text{NL} +125} \right) +b_\mathrm{l},
\end{equation}
for 
\begin{equation}
f_\text{NL}\geq -\frac{5}{4},
\end{equation}
with $a_q,~b_q,~c_q,~a_\mathrm{l},~b_\mathrm{l}$ fitting parameters (see Sec.\ \ref{sec:toy}) that might depend on the number of pixels used. Note that relations between the spectral index and the scalar amplitude of the primordial power spectrum can also be derived for other inflation models, e.g., the modulated Higgs inflation scenario of Sec.\ \ref{MHI}.

Analogously one could solve the linear fitting formula of the spectral index for $\kappa$ and substitute the latter within the quadratic fitting formula to derive a relation between $A_s$ and $n_s$. This yields 
\begin{equation}
10^9 A_s(n_s) = \frac{a_q}{a_\mathrm{l}^2}\left(n_s -1 - b_\mathrm{l} \right)^2 + \frac{b_q}{a_\mathrm{l}}\left(n_s -1 - b_\mathrm{l} \right) +c_q,
\end{equation}
whereby we have not neglected the contributions of the trispectrum.

\bibliography{bibliography}
\end{document}